\documentclass[12pt, preprint, longabstract]{aastex}
\usepackage{ccaption,equation}
\usepackage[multidot]{grffile}
\newcommand\ltsima{$\; \buildrel <\over\sim \;$}
\newcommand\simlt{\lower.5ex\hbox{\ltsima}}
\newcommand\gtsima{$\; \buildrel >\over\sim \;$}
\newcommand\simgt{\lower.5ex\hbox{\gtsima}}
\def\VEV#1{\left\langle #1\right\rangle}

\shorttitle{MOA-II bulge microlensing event rate and optical depth}
\shortauthors{Sumi et al.}

\begin{document}

\title{The Microlensing Event Rate and Optical Depth Toward the Galactic Bulge from MOA-II}

\author{
T.~Sumi\altaffilmark{1}, 
D.P.~Bennett\altaffilmark{2},
I.A.~Bond\altaffilmark{3},
F.~Abe\altaffilmark{4},
C.S.~Botzler\altaffilmark{5},
A.~Fukui\altaffilmark{6},
K.~Furusawa\altaffilmark{4},
Y.~Itow\altaffilmark{4},
C.H.~Ling\altaffilmark{3},
K.~Masuda\altaffilmark{4},
Y.~Matsubara\altaffilmark{4},
Y.~Muraki\altaffilmark{4},
K.~Ohnishi\altaffilmark{7},
N.~Rattenbury\altaffilmark{5},
To.~Saito\altaffilmark{8},
D.J.~Sullivan\altaffilmark{9},
D.~Suzuki\altaffilmark{1},
W.L.~Sweatman\altaffilmark{3},
P.,J.~Tristram\altaffilmark{10},
K.~Wada\altaffilmark{1},
P.C.M.~Yock\altaffilmark{5}\\ 
(The MOA Collaboratoin) \\
}

\altaffiltext{1}{Department of Earth and Space Science, Graduate School of Science, Osaka University, Toyonaka, Osaka 560-0043, Japan,\\
e-mail: {\tt sumi@ess.sci.osaka-u.ac.jp}}
\altaffiltext{2}
{Department of Physics, University of Notre Dame, Notre Dame, IN 46556, USA; bennett@nd.edu}
\altaffiltext{3}
{Institute of Information and Mathematical Sciences, Massey University,
Private Bag 102-904, North Shore Mail Centre, Auckland, New Zealand;
i.a.bond,c.h.ling,w.sweatman@massey.ac.nz}
\altaffiltext{4}{Solar-Terrestrial Environment Laboratory, Nagoya University, Nagoya 464-8601, Japan; abe,furusawa,itow,kmasuda,ymatsu@stelab.nagoya-u.ac.jp}
\altaffiltext{5}
{Department of Physics, University of Auckland, Private Bag 92019, Auckland, New Zealand; c.botzler,p.yock@auckland.ac.nz}
\altaffiltext{6}{Okayama Astrophysical Observatory, National Astronomical Observatory of Japan, 3037-5 Honjo, Kamogata, Asakuchi, Okayama 719-0232, Japan}
\altaffiltext{7}{Nagano National College of Technology, Nagano 381-8550, Japan}
\altaffiltext{8}{Tokyo Metropolitan College of Aeronautics, Tokyo 116-8523, Japan}
\altaffiltext{9}{School of Chemical and Physical Sciences, Victoria University, Wellington, New Zealand}
\altaffiltext{10}{Mt. John University Observatory, P.O. Box 56, Lake Tekapo 8770, New Zealand}

\begin{abstract}
We present measurements of the microlensing optical depth 
and event rate toward the Galactic Bulge based on two years of the MOA-II survey.  
This sample contains $\sim 1000$ microlensing events, with an Einstein Radius crossing time of
$t_{\rm E} \leq 200\,$days in 22 bulge fields covering $\sim42$ deg$^2$ 
between $-5^\circ <l< 10^\circ$ and $-7^\circ <b< -1^\circ$. Our event rate
and optical depth analysis uses 474 events with well defined microlensing parameters.
In the central fields with  $|l|< 5^\circ$,
we find an event rates of   
$\Gamma = [2.39\pm1.1]e^{[0.60\pm0.05](3-|b|)}\times 10^{-5}\,$star$^{-1}$\,yr$^{-1}$
and an optical depth 
(for events with $t_{\rm E}\leq 200\,$days) of 
$\tau_{200}  = [2.35\pm0.18]e^{[0.51\pm0.07](3-|b|)}\times 10^{-6}$ 
for the 427 events using all sources brighter than $I_s \leq 20$ mag. The distribution
of observed fields is centered at 
$(l,b)=(0.^\circ38, -3.^\circ72)$. 
We find that the event rate is maximized at low latitudes and a longitude of $l \approx 1^\circ$.
For the 111 events in $3.2\,{\rm deg}^2$ of
the central Galactic Bulge at $|b| \leq 3^\circ.0$ and $0^\circ.0 \leq l \leq 2^\circ.0$,
centered at  $(l,b)=(0.^\circ97, -2.^\circ26)$, we find 
$\Gamma =  4.57_{-0.46}^{+0.51}  \times 10^{-5}$\,star$^{-1}$\,${\rm yr}^{-1}$ and 
$\tau_{200} =  3.64_{ -0.45}^{+  0.51}  \times 10^{-6}$. 
We also consider a Red Clump Giant (RCG) star
sample with $I_s < 17.5$,
and we find that the event  rate for the RCG sample is slightly lower
than but consistent with the all-source event rate. The main difference is the
lack of long duration events in the RCG sample, due to a known selection
effect. Our results are consistent with previous optical depth measurements,
but they are somewhat lower than previous all-source measurements and
slightly higher than previous RCG optical depth measurements. This suggests
that the previously observed difference in optical depth measurements
between all-source and RCG samples may be largely due to statistical
fluctuations. 
These event rate measurements towards the central galactic bulge
are necessary to predict the microlensing event rate and to optimize
the survey fields in the future space mission such as WFIRST.
\end{abstract}


\section{Introduction}
The gravitational microlensing surveys toward the Galactic Bulge (GB)
have been shown to be useful for exoplanet searches, the study of the structure, 
kinematics and dynamics of the Galaxy, and measurement the 
stellar and sub-stellar mass functions, as the event rate and timescale distributions
are related to the masses and velocities of lens objects (\citealt{pac91}, \citealt{gri91}, \citealt{novati2008}).
To date, several thousands of microlensing events have been detected in the GB 
by the microlensing survey groups: OGLE (\citealt{uda94, uda00, woz01, uda03,sumi2006}), 
MOA (\citealt{bon01, sumi03}), MACHO (\citealt{alc97, alc00b}) and 
EROS (\citealt{afo03,ham06}). Thousands of detections are expected in the
upcoming years from the
MOA-II\footnotemark\footnotetext{\tt http://www.massey.ac.nz/\~{}iabond/alert/alert.html}, 
OGLE-IV\footnotemark\footnotetext{\tt http://www.astrouw.edu.pl/\~{}ogle/ogle4/ews/ews.html},
and
WiSE\footnotemark\footnotetext{\tt http://wise-obs.tau.ac.il/\~{}wingspan/} \citep{wise_survey}
surveys, which are currently in operation. These surveys will soon be joined by the 
KMTNet survey \citep{kmtnet}.

The magnification of a microlensing event is described by (\citealt{pac86})

\begin{equation}
  \label{eq:amp-u}
  A(u)= \frac{u^2+2}{u\sqrt{u^2+4}},
\end{equation}

\noindent
where $u$ is the projected separation of the source and lens 
in units of the Einstein radius $R_{\rm E}$ which
is given by

\begin{equation}
  R_{\rm E}(M,x) = \sqrt{\frac{4GM}{c^2}D_{\rm s}x(1-x)},
  \label{eq:re}
\end{equation}

\noindent
where $M$ is the lens mass, $x=D_{\rm l}/D_{\rm s}$ is the 
normalized lens distance and $D_{\rm l}$ and $D_{\rm s}$ are 
the observer-lens and the observer-source distances. 
The time variation of $u=u(t)$ is

\begin{equation}
  \label{eq:u}
  u(t)=\sqrt{u_{\rm 0}^2 + \left( \frac{t-t_{0}}{t_{\rm E}} \right)^2},
\end{equation}

\noindent
where $u_{\rm 0}$, $t_{0}$, $t_{\rm E}= R_{\rm E}/v_{\rm t}$ and 
$v_{\rm t}$ are, respectively, the minimum impact parameter in units of 
$R_{\rm E}$, the time of maximum magnification, the Einstein radius
crossing time (or timescale), 
and the transverse velocity of the lens relative to the line of sight 
toward the source star. From light curve alone, one can determine the 
values of $u_{\rm 0}$, $t_{0}$ and $t_{\rm E}$, but not the values 
of $M$, $x$ or $v_{\rm t}$.

The microlensing optical depth, $\tau$ is the fraction of the sky covered by the
Einstein ring disks of the lenses for a given source population, and
it is directly related to the mass density of compact objects along the 
line of sight (\citealt{pac96}). Theoretically, it is simpler than the 
microlensing event rate, $\Gamma$, because it doesn't depend on
the lens and source velocity distribution. Practically, it is difficult
to measure, however, because long duration events give a large
contribution to $\tau$, and it is difficult to ensure that there is not
a significant contribution from events with a duration longer than
the maximum $t_{\rm E}$ for a given analysis. Because of this, we present
observed values of the optical depth with a subscript, which indicates
the maximum $t_{\rm E}$ value allowed by the analysis of each observational
sample.

Despite this ambiguity due to long duration events, 
previous Galactic bulge microlensing optical depth results have been somewhat
controversial.  \cite{pac91} and \cite{gri91} first predicted the 
optical depth of $\tau\sim 5\times 10^{-7}$, assuming that all events 
were associated with known disk stars. After the first several bulge 
events were reported by OGLE (\citealt{uda94})
and MACHO \citep{alc95}, the high event rate 
prompted \cite{kir94} to evaluate the contribution of bulge stars in 
addition to the disk stars. They estimated $\tau\sim 8.5\times 10^{-7}$ 
and concluded that the value could be about twice as large, if the 
bulge were elongated along the line of sight. Nevertheless, the first
measurements of the optical depth, $\tau_{100}\sim3.3 \times 10^{-6}$ by 
OGLE (\citealt{uda94}) and $\tau_{150}\sim3.9^{+1.8}_{-1.2} \times 10^{-6}$ 
by MACHO (\citealt{alc97}), were well above the predictions.
The later studies based on Difference Image Analysis (DIA), which is less 
sensitive to the systematics of blending in crowded fields, also 
found relatively large optical depths: 
$\tau_{150}=2.43^{+0.39}_{-0.38} \times 10^{-6}$ centered 
at $(l,b)=(2.^{\circ}68, -3.^{\circ}35)$ from 99 events by MACHO (\citealt{alc00b}) 
and 
$\tau_{150} =2.59^{+0.84}_{-0.64} \times 10^{-6}$ centered
at $(l,b)=(3.^{\circ}0, -3.^{\circ}8)$ from 28 events by MOA
(\citealt{sumi03}) (where the dubious adjustment factor to
``correct" for foreground disk stars has not been uesed.) This MACHO DIA analysis
had one known $t_{\rm E} \sim 500$-day event that was removed by the $t_{\rm E} < 150\,$day
cut, and the inclusion of this single event would have raised the measured
$\tau$ value by $\sim 15$\%.

To explain high optical depths a number of authors have suggested the 
presence of a bar oriented along our line of sight to the GB 
(\citealt{pac94}; \citealt{zha95}), and have adopted various values of 
the bar orientation and mass 
(\citealt{han03}; \citealt{zha96}; \citealt{pea98}; \citealt{gyu99}). 
The resulting values are in the range $\tau= 0.8 - 2.0 \times 10^{-6}$. 
\cite{bin00} have shown that high optical depth measurements available
at the time could not be easily reconciled with our general understanding 
of the Galactic dynamics, and that the standard models of the Galaxy 
would need to be revised.

\cite{alc97} raised the possibility of a systematic bias in the 
optical depth measurement due to the difficulties of measuring 
$t_{\rm E}$ associated with blended unresolved sources. When 
the actual source base-line flux is unknown, $t_{\rm E}$ and 
$u_{\rm 0}$ are degenerate in relatively low signal-to-noise
ratio (S/N) events (c.f. \citealt{woz97}; \citealt{han99}; 
\citealt{bon01}; \citealt{gou02}). \cite{pop01}
postulated that optical depth may be estimated without a bias 
due to blending by using only events with bright source stars, 
such as red clump giants, in which they thought that 
the blending might be negligible, rather than using all stars 
including the faint sources as in previous studies.
Although the first measurements by \cite{alc97} gave a high value 
$\tau_{150} \sim3.9^{+1.8}_{-1.2} \times 10^{-6}$,  the later measurements 
based on events with bright sources have returned lower optical depths: 
$\tau_{400} = 0.94 \pm 0.29 \times 10^{-6}$ 
at $(l,b)=(2.^{\circ}5, -4.^{\circ}0)$ from 16 events by EROS (\citealt{afo03}),
$\tau_{350} = 2.17^{+0.47}_{-0.38} \times 10^{-6}$ at
$(l,b)=(1.^{\circ}50, -2.^{\circ}68)$ from 42 events by MACHO (\citealt{pop05}),
and $\tau_{400} = (1.62 \pm 0.23) \exp [-a(|b| - 3 {\rm deg})]  \times 10^{-6}$,
with $a = (0.43 \pm 0.16) {\rm deg}^{-1}$ based on 120 EROS events
\citep{ham06}. 

All the bright star analyses discussed above have one drawback 
compared to the DIA analyses of \citet{alc00b} and \citet{sumi03}. In each case, the
analysis implicitly assumes that the events that occur at the approximate location
of a bright star are due to lensing of the bright star and not some much fainter star
at almost the same location. Actually, \citet{pop05} and \citet{ham06} realized that
lensing of a fainter star that is unresolved from the bright star would be relatively
common. However, they presented arguments that this would increase the apparent
number of bright star events, but it would also lead to shorter apparent event
durations. They argued that these two effects would nearly cancel, so that
these apparently severe blending effects would not lead to a large bias in the
optical depth. A comparison with theory seemed to support this conclusion.
These values are consistent with predictions based on the revised COBE bar model 
by \cite{han95}, which has a mass of $M_{\rm bulge}=1.62\times 10^{10} M_\odot$ 
and the viewing angle $\phi\sim 20^\circ$,and the latest COBE elongated bar 
model by \cite{bis02} with $\phi\sim 20^\circ$.

Although the optical depth results from the bright RCG sources and the fainter
sources from the DIA analyses are generally consistent within their error bars, the
DIA optical depth values are about 25\% larger than the RCG values. The reason
for this is not well understood. One possibility is that the difference is simply 
statistical, and it is just a coincidence that the two all-source analyses have
given larger optical depths than the RCG analyses.
\cite{sumi2006} and \cite{smith2007} conducted detailed image level simulations on systematics
due to the blending, and found that the bright RCG samples do suffer from 
systematic errors and suffer biases although the biases do tend to cancel
at the level of the statistical uncertainties in previous work, in agreement
with \citet{pop05} and \citet{ham06}.
\cite{sumi2006} measured the Galactic bulge optical depth from
OGLE-II data for RCG sources with high signal-to-noise light curves, which allowed
the source brightness to be determined from the microlensing light curve fit. This
allowed the source brightness to be determined so that RCG ``impostor" events,
with faint sources, could be excluded. This yielded an optical depth value,
$\tau_{150} = 2.55_{-0.46}^{+0.57} \times 10^{-6}$ at $(l,b)=(1.^{\circ}16, -2.75^{\circ}8)$,
slightly higher than, but quite consistent with the MACHO and EROS values.
This value is also slightly lower than the analyses using DIA photometry of
events with fainter sources. This suggests that this difference in the optical depth
for the bright and faint star samples might be due to some physical effect,
as was seen in the models of \citet{kerins2009}. However, another possibility
is that there is a problem with estimating the number of sources as a function
of position in the bulge.

While the previous discussion has focused on the microlensing optical depth,
due to its simpler theoretical interpretation, the microlensing rate has 
a number of advantages. It is a more direct measure of the number of
microlensing events that will be seen by a future space-based microlensing
survey \citep{ben02}, like the exoplanet microlensing survey planned for
WFIRST \citep{green2012} or Euclid \citep{penny13}
The event rate has smaller uncertainties because it
is not dominated by a small number of very long duration events.

In this paper we present a measurement of the microlensing event rate and
optical depth toward the GB based on the first two years of the MOA-II survey,
the second phase of the MOA experiment. Our analysis makes use of
only high S/N events with well constrained model parameters to avoid the 
ambiguities due to parameter degeneracies caused by blending.
We present the photometric data in section \S\,\ref{sec:data} and the
selection of microlensing events in section \S\,\ref{sec:select}.
In section \S\,\ref{sec:eff}, we present the computation of the detection efficiency, and
we present the event rate and optical depth results in section \S\,\ref{sec:opt}. 
In section \S\,\ref{sec:model}, we discuss models of the variation of the
event rate and optical depth with galactic coordinates, and we discuss out
results and present our conclusions in section \S\,\ref{sec:summary},

\section{Data}
\label{sec:data}

The data set used in this analysis is same as used by \cite{sumi2011}. 
That is, it was taken in the
2006 and 2007 seasons by the MOA-II survey,
with the 1.8-m MOA-II telescope located at the Mt.\ John University
Observatory, New Zealand.
The telescope is equipped with the wide field camera, MOA-cam3, which consist of ten
2k $\times$ 4k pixel CCDs with $15\,\mu$m pixels. With the pixel scale
of 0.58 arcsec/pixel scale \citep{sako2008}, this gives a
2.18 deg$^2$ field of view (FOV).
The median seeing for this data set was $2.0''$. 

The MOA-II survey is a high cadence photometric survey of millions of stars in 22 GB fields.
The centers of these fields are listed in Table~\ref{tbl:fld}, although field 22 was
not used in this analysis, due to its distance from the bulge.
The data consists of $\sim 8250$ images of each of the two most densely sampled fields,
gb5 and gb9,  with a 10 minute sampling cadence, and 1660-2980 images 
of each of the 19 other fields, which were sampled with a 50 minute 
cadence, as indicated in Table \ref{tbl:fld}. 
This high cadence strategy is designed to detect very short events with 
$t_{\rm E} < 2$ days, which are expected due to lensing by
free-floating planets \citep{sumi2011} and short planetary anomalies
in the light curves of stellar microlensing events \citep{mao1991,sumi2010,bennett_rev,gaudi_rev}, 
This high cadence also increased the fraction of the events
in which the lensing parameters are well constrained by the light curve fit, which is 
very important for precise measurements of the microlensing rate and optical depth.

The images were reduced with MOA's implementation \citep{bon01} of 
the difference image analysis (DIA) method \citep{tom96,ala98,ala00}. 
In the DIA method, a high quality,
good seeing, reference image is subtracted from each observed
image after transforming the reference image to give it the same seeing
and photometric scaling as the observed image. This method generally
provides more precise photometry in the very crowded Galactic bulge fields
than PSF-fitting routines, such as DOPHOT \citep{sch93}.
This is, in part, due to the fact that the GB fields are
so crowded that virtually all the main sequence stars
are not individually resolved. As described in Section \ref{sec:eff},
the identification of a clear RCG population in the data
is needed to match the observed MOA luminosity function to the much
deeper Hubble Space Telescope (HST) luminosity function 
\citep{hol98} that describes the source stars. This is similar to the
OGLE-II optical depth analysis \citep{sumi2006} that makes use of the OGLE-II 
extinction map \citep{sumEX04}, which is based on the RCG position 
in the color magnitude diagram (CMD) of each field.

Each field is divided into 80 subfields and each subfield is individually
calibrated using the RCG feature in each subfield CMD.
For the gb22 field and some fraction of other fields, totaling
about 12\% of the area, a clear RCG population could not
be identified in the CMD, and these regions
were excluded from the analysis for this reason. 
The number of subfields used in the final analysis is 1536 in total and also 
given in Table~\ref{tbl:fld} for each field,
where the maximum is 79 as one subfield is not useful for a technical reason.
The coordinates and other information of subfields are listed in  Table~\ref{tbl:sub_eff_all}.
This lack of a clear RCG feature in field gb22 is because it
is relatively far from the center of the Galaxy. For the other fields,
It is generally regions of
very high interstellar extinction that prevented the identification of the
RCG CMD feature in some of the subfields.

The images were taken using the custom MOA-Red wide-band filter, which
is equivalent with the sum of the standard Kron/Cousins $R$ and $I$-bands.
The instrumental magnitudes of the MOA reference images were
calibrated to the Kron/Cousins $I$-band using OGLE-II photometry map of the
Galactic bulge \citep{uda02}.
The mean magnitude zero-point were estimated from the 30\% of MOA-II fields which
overlap with the OGLE-II map. We applied this mean zero-point to all fields.
The uncertainty in the magnitudes calibrated by this procedure is
estimated to be $\sim$0.25 mag
from the standard deviation of zero-points in overlap fields.
Although this calibration is approximate, it does not affect
following analysis at all because the luminosity functions, which is the
only part of our analysis requiring calibrated magnitudes, are
calibrated by using the RCG CMD feature, as discussed in Section \ref{sec:eff}.

\section{Microlensing event selection}
\label{sec:select}

We make use of the same microlensing events selected in the analysis
of \cite{sumi2011}.
The processes and criteria for event selection have been developed for that
paper and are reused for the current analysis. Event selection details
are summarized in Table S2 and Section 2 of the 
\citet{sumi2011} Supplementary Information (SI). 

In short, we have selected light curves with a single instantaneous 
brightening episode and a flat constant baseline, which can be well 
fit with a point-source, point-lens (PSPL) microlensing model given by Eq. (\ref{eq:amp-u}),
with separate parameters for both the source and blend fluxes.
We required that the lensing parameters are well constrained by the
light curve model, including both the source and blend fluxes, with
the minimum impact parameter of $u_0<1.0$.
Binary lens events are excluded from this analysis with a strict cut on 
the $\chi^2$ of the PSPL model. 
Although we have identified more than a thousand microlensing candidates
in this data set, only 474 high quality microlensing events have passed our
relatively strict cuts on the error bars of the event parameters as determined
by the microlensing model fit. These strict criteria ensure that $t_{\rm E}$ is well constrained
for each event and that there is no significant contamination by
mis-classified events. Thus, they ensure that the event rate and
optical depth measurements are not significantly biased by low-level
systematic errors in the event parameters.
The agreement of the observed and simulated $u_0$ distributions, shown
in Fig.~S4 in the SI of \cite{sumi2011} support
the conclusion that systematic errors or contamination by non-microlensing
variability are negligible. Fig.~S5 of the SI indicates that the 
systematic bias between input and the fit $t_{\rm E}$ of simulated
events is $\simlt 5$\% level regardless of $t_{\rm E}$ as seen in 
Fig.~S5 of SI in \cite{sumi2011}, which also support the negligible
bias in our optical depth estimates.

The number of selected events, $N_{\rm ev}$, in each field and subfield are listed in 
Tables~\ref{tbl:fld}, \ref{tbl:sub_eff_all} and  \ref{tbl:sub_eff_RCG}, respectively.
In Table \ref{tbl:candlist}, we list the microlensing events with the best 
fit parameters, with the complete list available in the online, electronic
edition, and a sample available in the print edition.

\subsection{Defining Source Populations}
\label{sec:level0}

\begin{figure}
\begin{center}
\includegraphics[angle=-90,scale=0.7,keepaspectratio]{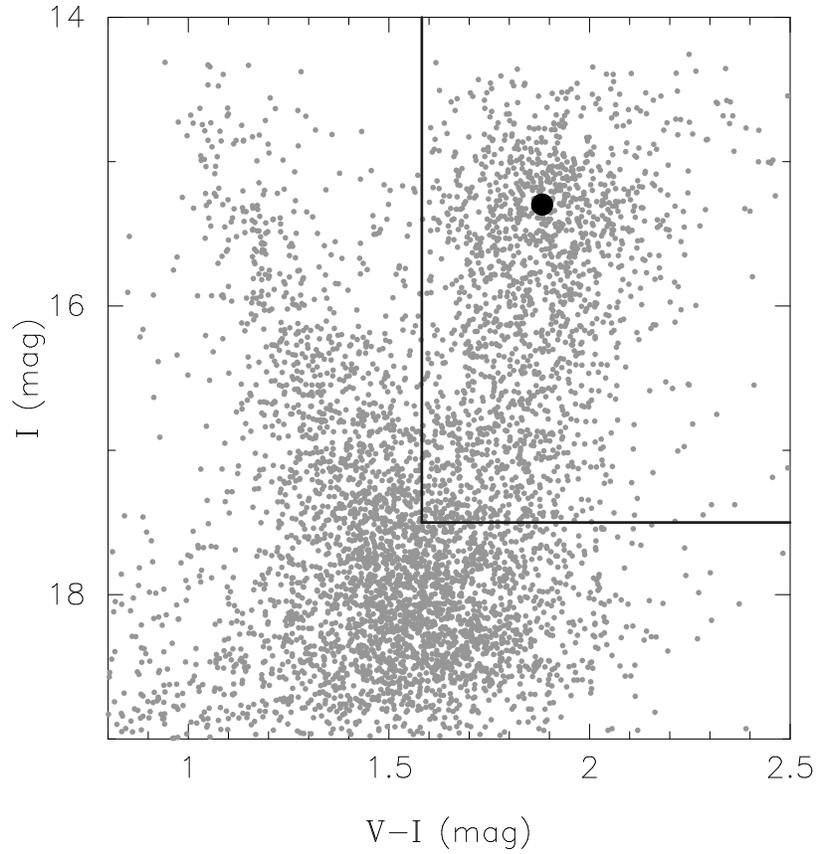}
\caption{
  \label{fig:CMD}
The ($V-I$, $I$) color magnitude diagram of a subfeild, gb13-8-3 at ($l,b$)=($2.^\circ10,  -4.^\circ05$).
The filled circle indicates the RCG centroid, $(V-I, I)_{\rm RC}$. The stars in  "extended RCG region"  
defined by the solid lines, i.e., $I<17.5$ mag, $V-I \geq (V-I)_{\rm RC}+0.3$ mag 
are used for our RCG sample.
}
\end{center}
\end{figure}

For the microlensing rate and optical depth estimates,
we use two subsamples of events:
(1) the all-sources sample, with events brighter than $I_s \leq 20$ mag, and 
(2) the Red Clump Giant (RCG) sample, the stars in the "extended RCG region"
as shown in Figure \ref{fig:CMD}, which is a similar definition as previous works.
This will allow us to see if there is a discrepancy as was seen in previous studies.

(1) The "all source" sample uses all 474 events mentioned above. 
In this sample, the events with the best fit source magnitude of 
fainter than $I_{\rm s}=20$ mag have been rejected to avoid the 
possible contamination from the events with degenerate parameters.

(2) The RCG sample uses 83 events with source magnitudes
of $I_{\rm s}<17.5$ mag and the source colors of $(V-I)_s\geq(V-I)_{\rm RC}-0.3$ mag,
where $(V-I)_{\rm RC}$ is the $V-I$ color of RCG centroid.
The RCG centroid ($V-I$, $I$)$_{\rm RC}$ was identified in each subfield
by a method very similar to that of  \cite{Nataf2012}.
This RCG sample is similar to the ones in the previous RCG microlensing
optical depth paper papers \citep{alc97,pop05,sumi2006,ham06},
which contain not only the bulge RCG, but also contain other bulge red giants 
in the "extended RCG region" of the CMD as shown in Figure  \ref{fig:CMD}.
Because we don't have $V$-band source magnitudes from the light curve models,
we use the $V$ and $I$ photometry of stars on the reference image
at the position of the event for the source color, $(V-I)_s$.
Our event detection method on the difference images does
not require the star to be associated with an apparently resolved 
reference image star, but we do require that fit source magnitudes are
$I_{\rm s}<17.5$ mag. So, in almost all cases, the bright star will be the source or
the source blended with another bright star. A fraction of these bright stars and events do 
not have $V-I$ photometry, mostly due to high extinction, which renders the stars
undetectable in the $V$-band. Since a blue star would have to be intrinsically very
bright to pass the $I_{\rm s}<17.5$ mag cut in these high extinction regions, these stars
are very likely to satisfy our color requirement, even though we are unable to 
verify that they do.
In practice, we have rejected only stars that have a clear measurement of 
$V-I$ with $(V-I)_s<(V-I)_{\rm RC}-0.3$. We have kept the rest of them, because 
the purpose of  color cut is to reduce the contamination from foreground disk stars 
as much as possible.
We have used this condition for both selecting events and counting source 
stars for our RCG sample.

\section{Detection efficiency}
\label{sec:eff}

We use the detection efficiency determined by \cite{sumi2011},
as described in Sections S4 of the SI of that paper, but we briefly
summarize our method here. The detection efficiency of our survey
was determined with Monte Carlo simulations following \citet{sumi03}.
Artificial microlensing events were added at random positions to the
observed images, using PSFs derived from nearby stars in each field.
The parameters of these artificial events were uniformly generated
at random in the following ranges for the impact parameter, $u_0$,
time of peak magnification, $t_0$, Einstein radius crossing time,
$t_{\rm E}$, and source magnitude, $I_s$: $0 \le u_0\le 1.5$,
$2453824 \le t_0\le 2454420$ JD, $0.1 \le t_{\rm E} \le 250\,$days, and
$14.25\le I_s \le 21.0$ mag. (The $t_0$ range is
the range of observations in this data set.)
Although we select only events with $I_{\rm s} < 20$ mag, we simulate stars
down to $I=21$ mag in our detection efficiency calculations to take account
the fact that some events with sources magnitudes that are fainter than 
$I=20$ mag can be selected, due to the fit uncertainty in the source magnitude
values. We simulated events with up to $u_0  \le 1.5$ for the same reason.
The source magnitudes were weighted by the combined Luminosity Function (LF)
from MOA-II and the {\it Hubble Space Telescope} (HST) \citep{hol98}. This uses
the MOA-II LF at a subfield of gb13-5-4 for bright stars and HST for faint stars down to $I = 24$ mag.
(Subfield gb13-5-4 is the one that contains the \citet{hol98} Baade's WIndow field.)
This combined LF is calibrated to the extinction and Galactic bulge distance for
each subfield using the position of the RCG centroid in each CMD, 
because RCG stars serve as a good standard candle (\citealt{kir97,sta00}).
Subfields where the RCG feature in the CMD could not be clearly identified
were not used in this analysis.

\begin{figure}
\begin{center}
\includegraphics[angle=-90,scale=0.7,keepaspectratio]{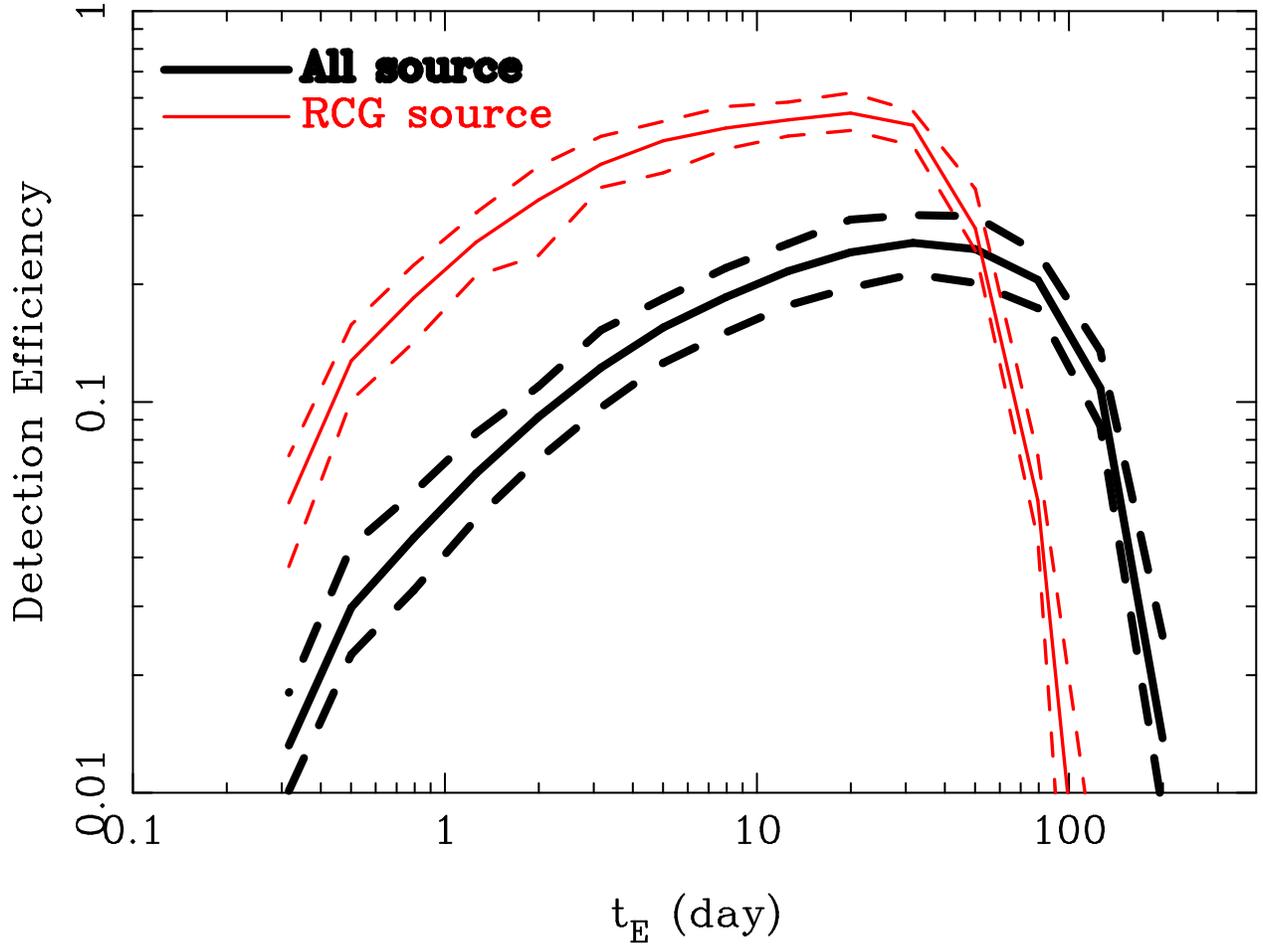}
\caption{
  \label{fig:eff}
The detection efficiencies of the MOA-II survey as a function of $t_{\rm E}$ 
for the all-source sample in think black and RCG sample in thin red. Solid and dashed lines 
indicate the mean, minimum and maximum efficiencies of all fields, respectively.
}
\end{center}
\end{figure}

Once the images with artificial events were created, they were processed with the
same analysis pipeline and selection criteria used for the analysis
of the actual data. We evaluated our detection efficiency as a function
of $t_{\rm E}$, $\varepsilon (t_{{\rm E}})$, in each field by simulating 20 million artificial events as shown in
Figure \ref{fig:eff}. For Einstein radius crossing times of 1-50 days, the
RCG efficiency is higher, as one would expect, but the RCG efficiency drops
steeply for events with $t_{\rm E}\simgt 80\,$days. This is an artifact of the 
original purpose of this analysis, which was to study the short timescale 
tail of the $t_{\rm E}$ distribution in order to search for a population of
free-floating planets \citep{sumi2011}. The original event selection
procedure included a cut on the $\chi^2$ for a constant brightness fit
for data outside of a 120-day window centered on the peak of the
event. Long duration events with bright sources are much more likely
to fail this cut.

In addition to the detection efficiency for each field as a function of the
event timescale, $\varepsilon (t_{{\rm E}})$, we have also determined the 
detection efficiency, averaged over $t_{\rm E}$, in each subfield, 
\begin{equation}
  \label{eq:epsilon}
<\varepsilon>=\frac{ \sum_i [\Gamma(t_{{\rm E},i}) \varepsilon (t_{{\rm E},i})]}{ \sum_i \Gamma(t_{{\rm E},i})},
\end{equation} 
where $\Gamma(t_{\rm E})$ is the $t_{\rm E}$ distribution appropriate for each subfield. 
Because we do not know the true $t_{\rm E}$ distribution for each subfield, 
we use a Gaussian weighted average of the 
observed detection efficiency corrected $t_{\rm E}$ distribution 
for all the other subfields within 1 degree around the subfield in question. The
Gaussian weighting uses $\sigma=0^\circ.4$ for the all-star sample, and
for the RCG sample, we use $\sigma=1^\circ$ for all sub-fields out to
$2^\circ.5$ from the subfield in question. The average efficiencies for each
subfield are listed in Tables \ref{tbl:sub_eff_all} and  \ref{tbl:sub_eff_RCG}
for the all-star and RCG samples, respectively.
These $<\varepsilon>$ will be used for analysis of event rates in the next section.
The average timescales for all source with 1 degree radius and  $\sigma=0.4$ degree 
are also shown in these tables and in the first panel of Figure \ref{fig:fieldmaps}.

\section{Microlensing Event Rate and Optical Depth}
\label{sec:opt}

The microlensing event rate, $\Gamma$, can be determined observationally
from the following expression,
\begin{equation}
  \label{eq:Gamma}
  \Gamma = \frac{1}{N_{\rm s}T_{\rm o}} \sum_i \frac{1}{\varepsilon (t_{{\rm E},i})},
\end{equation}
where $N_{\rm s}$ is the total number of source stars monitored for microlensing, 
$T_{\rm o}$ is the duration
of the survey in days, $t_{{\rm E},i}$ is the Einstein radius crossing time for 
the $i$-th event, and $\varepsilon (t_{{\rm E},i})$
is the detection efficiency at that time-scale.
The optical depth, $\tau$, is the probability that a random 
star is microlensed with the impact parameter $u_{\rm 0} \le 1$ at any 
given time. This is equivalent to the fraction of the total observing time
per star that a lens is within the angular Einstein radius of one of the source
stars. This depends on the impact parameter, $u_{\rm 0}$, but since the
distribution in $u_{\rm 0}$ is uniform, we can use the average event duration
(defined as the time when $u < 1$), which is given by $(\pi/2) t_E$.
We can substitute this average event duration into equation~\ref{eq:Gamma} to 
obtain the following expression for $\tau$, 
\begin{equation}
  \label{eq:opt}
  \tau = \frac{\pi}{2N_{\rm s}T_{\rm o}} \sum_i \frac{t_{{\rm E},i}}{\varepsilon (t_{{\rm E},i})}.
\end{equation}

A potential problem with equations~(\ref{eq:Gamma}) and (\ref{eq:opt}) occurs if 
$\varepsilon (t_{{\rm E}}) \rightarrow 0$ for some $t_E$ values. In this case, the 
uncertainty in $\Gamma$ or $\tau$ can diverge simply because we can't
detect events for some range in $t_{\rm E}$. This problem can be corrected
if we have some knowledge of the $t_E$ distribution. In practice, the main
difficulty occurs when trying to measure the optical depth, where a substantial
fraction of the total optical depth can come from very long duration events.
For the optical depth, the uncertainty is exacerbated by the large weight given to
events with large $t_{\rm E}$.
Some bulge events have been observed with 
durations of $t_{\rm E}\sim 500\,$days or more \citep{poi05}, so we add a subscript the
measured values of $\tau$ to indicate the maximum duration
of the events in the sample.

 In our event rate and optical depth analyses for this 2006-2007 data set, 
 $T_{\rm o}=596.0$ days and the number of source stars is 
(1) $N_{\rm *}=90.4\times 10^6$ for all star sample and  
(2) $N_{\rm *,RC}=6.49\times 10^6$ for the RCG sample.
These numbers were determined as follows:

(1) For the all star case,  we estimated the center of RCG $I$-band magnitude, $I_{\rm RC}$, 
and the number of RCG, $N_{\rm RC}$ by fitting the Luminosity function of the reference 
images in each subfield with Equation (4) of \cite{Nataf2012}.
The combined LF, which is based on LF in Baade's window field (gb13-5-4), 
are scaled and shifted so that its $I_{\rm RC}$ and $N_{\rm RC}$ are same as the 
values in each subfield. Then the number of stars $N_{\rm s}$ are counted down to $I=20$ mag
by this scaled-combined LF and shown in Table~\ref{tbl:fld} and \ref{tbl:sub_eff_all}.
The advantage of this method is that $N_{\rm RC}$ is less affected by the blending as
they are bright and the shape of LF around this range is roughly symmetric. The disadvantage is
that it assumed the LF in all fields are same as that of Baade's window \citep{hol98}.

(2) For the RCG case, we counted the the number of stars in the reference images 
in the extended RCG region with $I<17.5$ mag and $(V-I) > (V-I)_{\rm RC}-0.3$ mag, $N_{\rm s,RC}$.  
They are shown in Table \ref{tbl:fld} and \ref{tbl:sub_eff_RCG}.
The advantage of this method is that 
we are not assuming that the LF in all fields are same as
that of Baade's window. The disadvantage is that, as discussed in \cite{sumi2006},
the number of sources that can be lensed is slightly overestimated due to blending. 
The blending makes the stars to be looked brighter overall, however more fainter stars
are coming into the the magnitude range than brighter stars going out because the LF 
is the increasing function towards the fainter stars at $I\sim 17.5$ mag.
\cite{smith2007} estimated  $N_{\rm s,RC}$  is overestimated by $\sim 10$\% by 
image level simulations. We rescaled $N_{\rm s,RC}$ down by $10$\%.

Due to our event selection criteria (e.g., small $\chi^2$/d.o.f. for PSPL fit),
all events with significant binary lens features
were removed from the sample of event used to determine the event
rate and optical depth.
The fraction of binary lens events among all microlensing events 
has been estimated at 8\% (\citealt{jar02}),
6\% (\citealt{alc00a}), 3\% (\citealt{jar04}), 6\% (\citealt{sumi2006}). 
We use 6\% to correct our optical depth measurement for binary lens events 
excluded from the sample.  The event rate can be corrected by the ratio between 
the number of events with binary lenses and single lenses is $0.06/(1-0.06)$. 
Assuming that the lens system consists of two stars 
having the same typical time scale, the optical depth contribution of a binary 
lens event is $2^{1/2}$ times that of a single lens event.
It follows that the optical depth values and their errors have to be rescaled
by a factor 1.09.

\begin{figure}
\begin{center}
\includegraphics[angle=-90,scale=0.65,keepaspectratio]{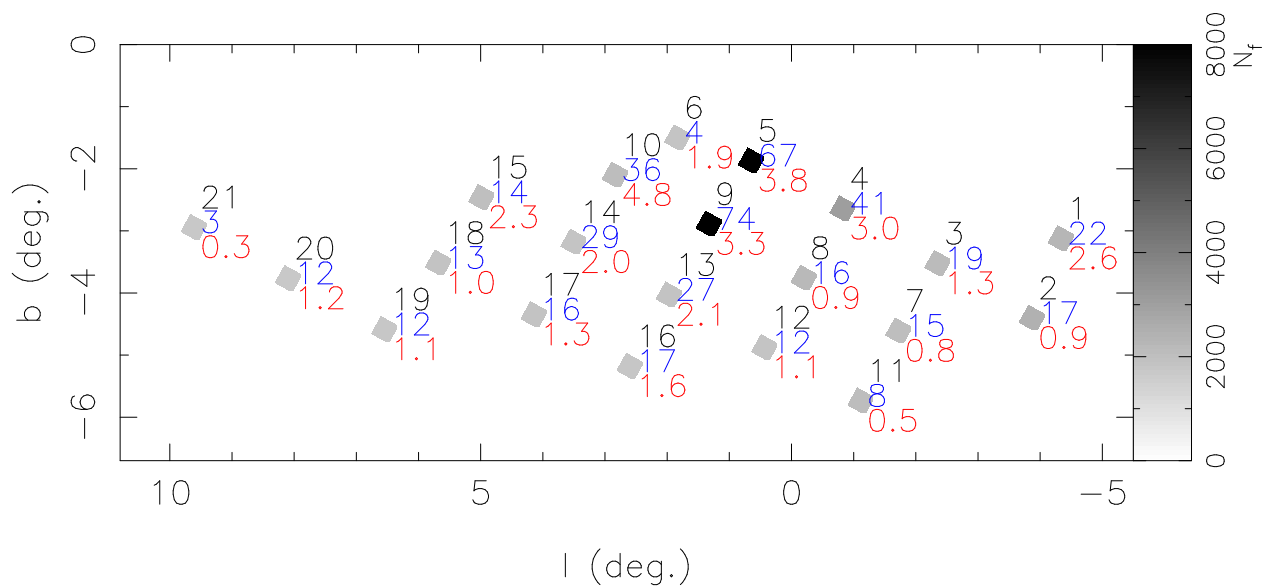}
\caption{
  \label{fig:field}
MOA-II galactic bulge fields used in this analysis.
The central galactic coordinates of fields are indicated by  diamonds
with field numbers (in black), the number of events (blue) and the optical 
depth $\tau_{200}\times 10^6$ (red), from the top to the bottom,
for the all-source sample. The color (or gray scale in the printed version) of the diamonds indicates the 
number of images per field, $N_{\rm f}$, as indicated by the scale on the right.
 }
\end{center}
\end{figure}

Individual optical depth estimates for all sources in each field 
are listed in Table~\ref{tbl:fld}, and shown in Figure~\ref{fig:field}.
The second panel of Figure~\ref{fig:fieldmaps} shows a smoothed version of the
optical depth centered at the location of each subfield.
We also estimated the average optical depth in all fields combined, and found
$\tau_{200}= 1.87_{ -0.13}^{+  0.15}\times 10^{-6}$ with 474 events for all source sample and
$\tau_{200}= 1.58_{-0.23}^{+0.27}\times 10^{-6}$ with 83 events for RCG sample 
at $(l,b)=(1.^\circ85, -3.^\circ69)$. The effective line of sight was 
computed by weighting the number of subfields used. The errors were 
estimated using the bootstrap Monte-Carlo method of \cite{alc97}. 

\subsection{Fitting with Poisson Statistics}
\label{sec:poisson_fit}

The event rate formula given in equation~(\ref{eq:Gamma}) has the apparent 
advantage that it does not have any dependence on the event $t_{\rm E}$ 
distribution to estimate the $\Gamma$. But this is a bit of an illusion, 
because we do need to know
something about the $t_{\rm E}$ distribution to estimate the uncertainty in $\Gamma$.
 With the bootstrap Monte-Carlo
method of \cite{alc97}, we assume that the true $t_{\rm E}$ distribution is just
the observed $t_{\rm E}$ distribution corrected for detection efficiencies. But, if
we do know the $t_{\rm E}$ distribution, then we can use a simpler formula for
$\Gamma$,
\begin{equation}
  \label{eq:Gamma-effav}
  \Gamma = \frac{1}{N_{\rm s}T_{\rm o}} \frac{N_{\rm ev}}{\langle\varepsilon \rangle}\ ,
\end{equation}
where $N_{\rm ev}$ is the number of events in the sample, and
$\langle \varepsilon \rangle$, 
is the detection efficiency, $\varepsilon (t_{\rm E})$, averaged over the assumed
$t_{\rm E}$ distribution, $\Gamma(t_{\rm E})$ appropriate for each subfield as defined 
by Equation~(\ref{eq:epsilon}).
If we assume the observed $t_{\rm E}$ distribution as $\Gamma(t_{\rm E})$, then we have
relations, 
$\sum_i \Gamma(t_{{\rm E},i}) = \sum_i [1/\varepsilon (t_{{\rm E},i})]/(N_{\rm s}T_{\rm o})$,
i.e., Equation (\ref{eq:Gamma}). By substituting this relation, Equation~(\ref{eq:epsilon}) becomes,

\begin{equation}
  \label{eq:mean_epsilon}
\langle\varepsilon\rangle
=\frac{ \sum_i \left[\frac{\varepsilon (t_{{\rm E},i})}{\varepsilon'(t_{{\rm E},i})} \right]}{ \sum_i    \frac{1}{ \varepsilon' (t_{{\rm E},i})}}
\simeq \frac{ N'_{\rm ev} }{ \sum_i    \frac{1}{ \varepsilon' (t_{{\rm E},i})}}
=\langle \frac{1}{ \varepsilon'}\rangle^{-1},
\end{equation}
where $N'_{\rm ev}$ and  $\varepsilon'$ are the number of events and detection 
efficiency in the area constructing the $t_{\rm E}$ distribution. The "$\simeq$" in the 
above equation is because we would construct the $t_{\rm E}$ distribution in 
the area where we can assume $\varepsilon' \simeq \varepsilon$.
So, Equation (\ref{eq:Gamma-effav}) does not remove the issue of uncertainties due to
low-efficiency events, which still affect the calculation of $\langle\varepsilon\rangle$
in Equation~(\ref{eq:mean_epsilon}).
However the advantage of Equation (\ref{eq:Gamma-effav}) is that we have separated 
"$N_{\rm ev}$" and  $ \langle \varepsilon  \rangle$ so that we can choice the larger 
sample around the subfield in question to minimize the statistical error of 
 $ \langle \varepsilon  \rangle$ even if there is only a few events in the subfield.
On the other hand, the traditional method by Equations (\ref{eq:Gamma}) imply that
only the detection efficiency of the detected events in the subfield in question are used.
So we have to enlarge the area of the subfield itself to reduce the statistical uncertainty
on $t_{\rm E}$ distribution, by which the signal should also be smoothed out.

Equation~(\ref{eq:Gamma-effav}) also has the advantage over
equation~(\ref{eq:Gamma}) that it obeys Poisson statistics, whereas equation~(\ref{eq:Gamma}) 
implies an unequal weighting of the different events. We therefore use
equation~(\ref{eq:Gamma}) to determine the $t_{\rm E}$ distribution, $\Gamma(t_{\rm E})$, 
within 1 degree around the subfield in question and then use this $t_{\rm E}$ distribution
to calculate $\langle \varepsilon \rangle$. 
This allows us to use equation~(\ref{eq:Gamma-effav})
to calculate $\Gamma$ for subsamples of the data.

The event rate, $\Gamma$, and optical depth, $\tau$, are expected to be continuous 
functions of $l$ and $b$. Attempts to measure the $l$ and $b$ dependence of 
have generally involved averaging $\Gamma$ or $\tau$ into bins and then
fitting functions of $l$ and $b$ to the binned data \citep{pop05,sumi2006,ham06}.
There are two problems with this procedure, however. First, the binning necessarily
smooths out the intrinsic spacial distribution in $l$ and $b$. Second, individual bins
often have a small number of events, but they are nevertheless fit using Gaussian
statistics, which do not apply. This leads to large fitting errors if there are bins with
0, 1, or 2 events.

In order to avoid these problems, we introduce a method for fitting with Poisson 
statistics, and this allows us to fit to the raw, subfield data, even though the average
number of events per subfield is $<1$. If the event rate predicted by a given model
for a subfield at coordinates $(l,b)$ is denoted by $\Gamma_{\rm mod}(l,b)$, then
from equation~(\ref{eq:Gamma-effav}), the number of expected events in that subfield is given by
\begin{equation}
  \label{eq:Nevexp}
  N_{\rm ev, exp}(l,b)=\Gamma_{\rm mod}(l,b) N_{\rm s}(l,b)T_{\rm o} \langle\varepsilon(l,b)\rangle \ ,
\end{equation}
where $N_{\rm s}$ is the number of stars in the subfield, $T_{\rm o}$ is the survey duration,
and $\langle\varepsilon(l,b)\rangle$ is the detection efficiency averaged over $t_{\rm E}$ for
the subfield at coordinates $(l,b)$. These are 
given in Tables \ref{tbl:sub_eff_all} and  \ref{tbl:sub_eff_RCG}, for the
all-star and RCG samples, respectively. 

The probability of the observed number of events, $N_{\rm ev}(l,b)$ ,in the subfield at $(l,b)$ is
\begin{equation}
  \label{eq:Poisson}
  P[N_{\rm ev}(l,b)]= \frac{ e^{-N_{\rm ev, exp}(l,b)} N_{\rm ev, exp}(l,b)^{N_{\rm ev}(l,b)}}{N_{\rm ev}(l,b)!} \ ,
\end{equation}
according to Poisson statistics. We can then define $\chi^2 = -2 \ln P[N_{\rm ev}(l,b)]$ for
each subfield, which implies that the $\chi^2$ for the full fit is
\begin{equation}
  \label{eq:Poisson-chi2}
  \chi^2 = -2 \sum_{(l,b)} \ln P[N_{\rm ev}(l,b)] = -2 \sum_{(l,b)} 
  \ln\left[  \frac{ e^{-N_{\rm ev, exp}(l,b)} N_{\rm ev, exp}(l,b)^{N_{\rm ev}(l,b)}}{N_{\rm ev}(l,b)!}\right] \ .
\end{equation}
This gives the $\chi^2$ value corresponding to the Poisson probability of the observed
number of event. This $\chi^2$ should generally behave like the more usual $\chi^2$
from Gaussian statistics, but there is one difference. If $N_{\rm ev, exp}(l,b)\approx 1$ for
a large fraction of the subfields, then we expect the $\chi^2$ per degree of freedom
to be $\chi^2/{\rm d.o.f.}\approx 2$. For $N_{\rm ev, exp}(l,b) = 1$, $\chi^2 = 2$ is the lowest
possible value (obtained for $N_{\rm ev}(l,b) = 0$, or 1). Thus, $\chi^2/{\rm d.o.f.}\approx 1$
is impossible when the number of events per subfield is close to 1.
We determine the fit parameter uncertainties by evaluating 68\%
confidence level with the Markov Chain Monte Carlo (MCMC).

In Section~\ref{sec:model}, we use this Poisson statistics fitting method for modeling
the event rate, $\Gamma$, but the modeling of the optical depth, $\tau$, is more complicated
because of the unequal weighting of events in equation~\ref{eq:opt}. Each event is summed
with a weight of $t_{{\rm E},i}/\varepsilon (t_{{\rm E},i})$, so the optical depth does not
follow Poisson statistics. Therefore, we use the standard binning method to model
the optical depth, $\tau$, distribution, but we take care to ensure that none of the bins
have a small enough number of events to invalidate the use of Gaussian statistics.

\section{Modeling the Event Rate and Optical Depth Results}
\label{sec:model}

Figure~\ref{fig:optball} shows the optical depth, $\tau_{200}$, as a function of $b$
for both the all-star and RCG samples. We show
 the results for the central region with $|l|<5^\circ$ to improve the overlap with previous measurements,
which are shown in the same figure.
The subfield results have been binned in bins of width $\Delta b= 0.5^{\circ}$.
The values for all-source and RCG  sample are listed in 
Tables \ref{tbl:opt_binb_all_lth5} and  \ref{tbl:opt_binb_RCG_lth5}, 
respectively.

The optical depth clearly increases with decreasing $|b|$, and a
simple exponential fit gives,
$\tau_{200}= [2.35\pm 0.18]\times10^{-6} \exp[(0.51\pm 0.07)(3-|b|)]$ for 
the all-source sample as indicated 
by the black solid line in Figure~\ref{fig:optball}. This result is slightly smaller than
the measurements by MOA-I \citep{sumi03} and MACHO \citep{alc00b} 
with all-source samples. It seems more consistent with 
the RCG measurements by MACHO \citep{pop05},
EROS-2 (\citealt{ham06}) and OGLE-II \citep{sumi2006} at $b\sim -3.5^\circ$, 
but somewhat higher at the lower latitude.
The exponential model seems to represent the data reasonably well in explaining the
optical depth measurements with the all-source sample.

An exponential fit for the optical depth toward RCG sources gives
$\tau_{200}= [1.64\pm 0.27]\times10^{-6} \exp[(0.47\pm 0.17)(3-|b|)]$,
which is indicated by the red solid line in the Figure~\ref{fig:optball}. 
This result is 
consistent with previous RCG measurements. The best linear fit to the
OGLE-II RCG measurements is indicated by the red dashed line 
in Figure~\ref{fig:optball}. 

The treatment of blending in the MACHO \citep{pop05} and EROS \citep{ham06}
analyses is rather crude. These analyses identify microlensing events 
solely by their proximity to apparent RCG stars identified in the reference
images, with no attempt to determine if the source is a RCG star or a fainter
main sequence star. These blending
effects will shrink apparent $t_{\rm E}$ values for all events, while increasing
the number of apparent RCG events. \citet{pop05} and \citep{ham06} 
make arguments to suggest that these two effects approximately cancel.
It is plausible that this cancelation still work in the accuracy presented 
in this analysis.

Note that the $\tau_{200}$ values for the all-source sample at  
$-3^\circ \simgt b \simgt -4^\circ$, slightly north Baade's window in field gb13, 
are consistent with the values for the RCG sample.
However $\tau_{200}$ is significantly 
higher for the all-source sample than for the RCG sample at $b>-3^\circ$.

Our new optical depth values for the RCG sample are consistent with the
previous RCG measurements \citep{pop05,sumi2006,ham06} and some older bulge models 
\citep{bis02,han03} and do
agree with the more recent model of \citet{kerins2009}

The uncertainty in $\tau$ is dominated by small number of long $t_{\rm  E}$ events.
For this particular analysis, this effect is exacerbated by the low efficiency for
events with $t_{\rm  E} > 100\,$days, due to the fact that the analysis was
originally designed to focus on short timesscale events. The high weight for
these long timescale events also means that the error bars for $\tau$ are difficult
to measure. Also, we directly measure $\tau_{200}$ instead of $\tau$, so a comparison
to Galactic models requires a correction based on the rate of very long events, which
we don't measure.

The event rate, $\Gamma$, does not have these problems, and so it is the preferred
quantity to compare to Galactic models, despite the simple theoretical interpretation
of the optical depth, $\tau$.
We show the event rate per star per year $\Gamma$ and the exponential fits for the 
all-source and RCG samples as a function of the galactic latitude, $b$, for $|l|<5^\circ$
in Figure \ref{fig:Gamma_vs_b} and in Tables~\ref{tbl:opt_binb_all_lth5}
and \ref{tbl:opt_binb_RCG_lth5}, respectively.  
The event rate has much less scatter
than $\tau_{200}$ and $\Gamma$ for both the all-source and RCG samples
are well fit by a simple exponential model. Note that these fits are done to the
subfield data using our Poisson statistics method (see Section \ref{sec:poisson_fit}), 
while the plots are binned for display in Figure \ref{fig:Gamma_vs_b} and in 
Tables~\ref{tbl:opt_binb_all_lth5} and \ref{tbl:opt_binb_RCG_lth5}.

The exponential model for the all-source and RCG samples are quite
similar with
$\Gamma_{\rm all}= [23.92\pm 1.13]\times10^{-6} \exp[(0.60\pm 0.05)(3-|b|)] {\rm yr}^{-1}$ 
per star for the all-source sample and
$\Gamma_{\rm RC}= [21.86\pm 2.67]\times10^{-6} \exp[(0.65\pm 0.11)(3-|b|)] {\rm yr}^{-1}$
per star for the RCG sample. The RCG event rate is slightly smaller, 
but consistent with the all-source
event rate. The RCG slope is $0.4\sigma$ steeper and the amplitude is
is 8\% or $0.7\sigma$ smaller. If we use the all-source model parameters
for the RCG sample, $\chi^2$ increases by only $\Delta\chi^2 = 0.33$.

There may be some uncertainty in the true value of $\Gamma_{\rm all}$, due to the 
uncertainty in the luminosity function. We use the {\it HST} luminosity function measured
in Baade's window \citet{hol98}, but the true luminosity function could have
some dependence on Galactic coordinates. Fortunately,
this uncertainty is largely removed for microlensing survey
simulations if the same  \citet{hol98} luminosity function is used to estimate
the source star counts, as the uncertainty in the luminosity function would
largely cancel out.

The need to assume an luminosity function can be avoided entirely be considering
the event rate per square degree per year, $\Gamma_{\rm deg^2}$,
for source stars above a given magnitude threshold, which is $I_s \leq 20$ in
our case. This quantity
is, therefore, more directly determined by the observations. However, unlike
$\Gamma$ and $\tau_{200}$, $\Gamma_{\rm deg^2}$ does not depend
solely on the distribution of stars and lens objects in the Galaxy. (The lens objects
consist of stars, brown dwarfs, planets and stellar remnants.) The event rate
per square degree per year, $\Gamma_{\rm deg^2}$, also depends on the foreground
extinction. In the Galactic plane, $\Gamma$ should be maximized but
$\Gamma_{\rm deg^2}$ could drop to zero if the extinction is so high that no stars
are brighter than the magnitude threshold. Figure~\ref{fig:Gamma_sq_vs_b} which 
shows $\Gamma_{\rm deg^2}$ as a function of $b$ and an exponential fit to the data.
The effect of extinction can be seen in
the lowest $|b|$ bin, where $\Gamma_{\rm deg^2}$ drops to well below the fit.
As with the $\Gamma$ fits, the fitting
was done to the raw subfield data, and the data is binned for display only.
The binned $\Gamma_{\rm deg^2}$ values are also given in Table~\ref{tbl:opt_binb_all_lth5}.

\begin{figure*}
\begin{center}
\includegraphics[angle=-90,scale=0.65,keepaspectratio]{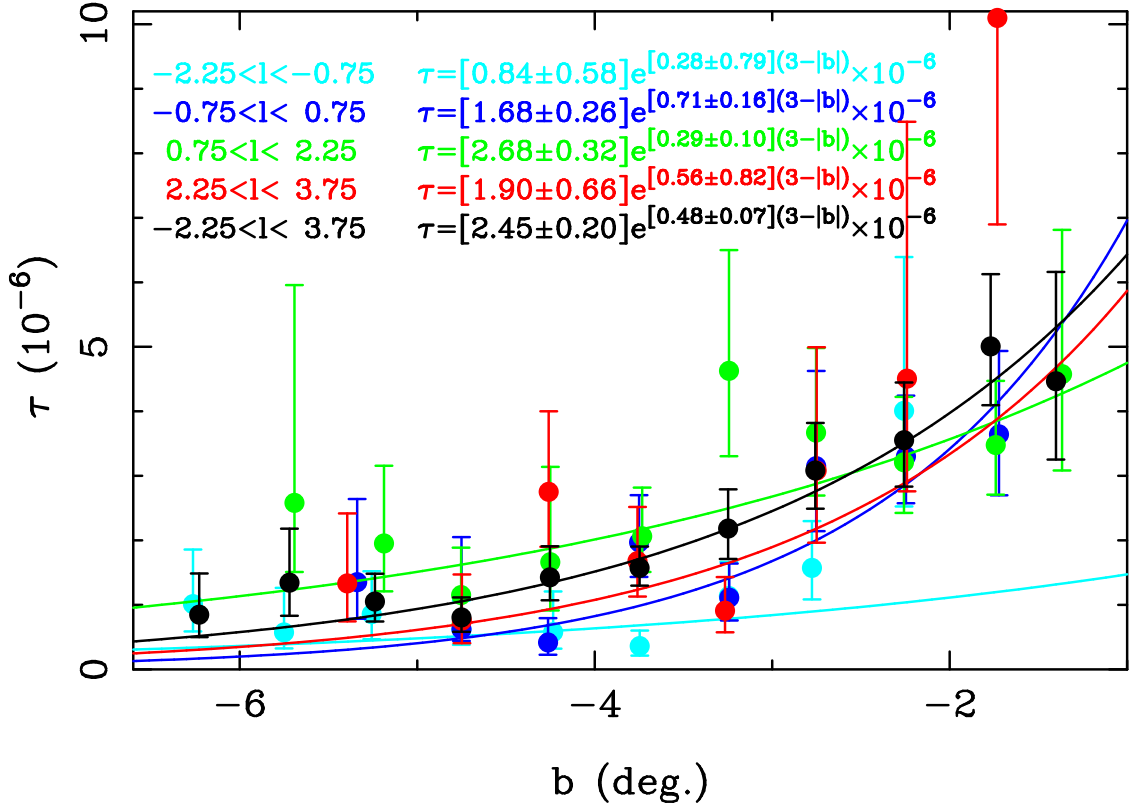}
\caption{The optical depth for events with $t_{\rm E} < 200\,$days, 
$\tau_{200}$, for the all-source sample 
as a function of the galactic latitude $b$ for different bins in Galactic
longitude, $l$. The curves show the best exponential fit in $b$.
The black curve is the fit to all the events with
$-2^\circ.25 < l < 3^\circ.75$, and it provides a reasonable fit to all the
longitude bins, except the $0^\circ.75 < l < 2^\circ.25$ bin,
where there is an enhancement to the rate.
\label{fig:tau_vs_b_l} 
}
\end{center}
\end{figure*}

\begin{figure*}
\begin{center}
\includegraphics[angle=-90,scale=0.65,keepaspectratio]{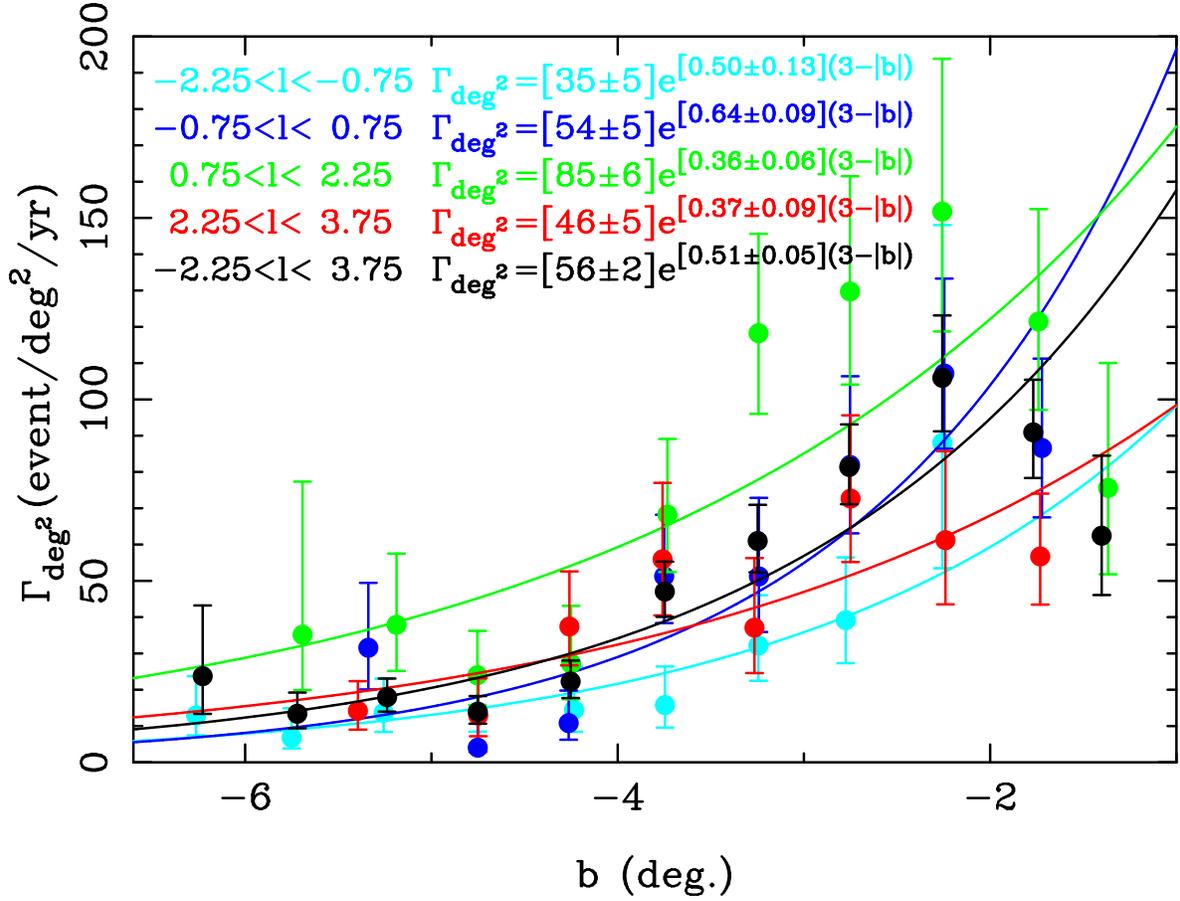}
\caption{The event rate per square degree per year, $\Gamma_{\rm deg^2}$, for the all-source sample 
as a function of the galactic latitude $b$ for different bins in Galactic
longitude, $l$. The curves show the best exponential fit in $b$ to the
unbinned subfield data.
The black curve is the fit to all the events with
$-2^\circ.25 < l < 3^\circ.75$.
The decline at $l>-2^\circ$ is due to the high extinction.
\label{fig:Gammasq_vs_b_l} 
}
\end{center}
\end{figure*}

We show exponential fits as a function of the galactic latitude $b$ for 
$\tau_{200}$, $\Gamma$ and $\Gamma_{\rm deg^2}$ for different bins 
in Galactic longitude, $l$ in Figures
\ref{fig:tau_vs_b_l},
\ref{fig:Gamma_vs_b_l} and
\ref{fig:Gammasq_vs_b_l}, respectively.
The black plots and curves are for all the events with
$-2^\circ.25 < l < 3^\circ.75$, and it provides a reasonable fit to all the
longitude bins, except the $0^\circ.75 < l < 2^\circ.25$ bin,
where there is an enhancement to the rate.

In the second through fourth panels of Figure \ref{fig:fieldmaps}, we display smoothed maps of
$\tau_{200}$, $\Gamma$, and $\Gamma_{\rm deg^2}$ as a function of Galactic
coordinates. The plotted values from all subfields are listed in Table \ref{tbl:opt_2D} of
the online version, with a sample of this table listed in the printed version of this paper.
The smoothing is done with a Gaussian function with $\sigma=0.4^\circ$, and cut off
at a distance of $1^\circ$ from the center of each subfield.
The error bars for each subfield are estimated using a bootstrap method using the
neighboring subfields with the same weighting as in the calculation of the central values.
As previously found by \citet{alc97} and \citep{pop05}, the highest optical depth is found
at $l\approx3^\circ$. This is due to the excess of long timescale events at this longitude,
as indicted in the first panel of Figure \ref{fig:fieldmaps}. The longitude of this
optical depth maximum is at the same longitude as MACHO field 104, which was noted
to have a excess of long timescale events by \citet{alc97} and \citep{pop05}, but we
we see $\tau_{200}$ maxima at $b\approx -2^\circ$ and $b\approx -5^\circ.5$, whereas MACHO
field 104 is centered at $b = -3^\circ.1$. However, this can probably be explained if there
is a real excess of long duration events at $l\approx 3^\circ$ at a range of 
latitudes. The differences between the MACHO and MOA results can probably be
explained by statistical fluctuations and uneven sampling of the $l\approx 3^\circ$ fields
by MACHO.

Both the event rate per star, $\Gamma$, and per square degree, $\Gamma_{\rm deg^2}$,
have a peak at $l\approx 1^\circ$. Because these event rate measurements obey
Poisson statistics, the statistical uncertainty in $\Gamma$ and $\Gamma_{\rm deg^2}$
is smaller than the uncertainty in $\tau_{200}$. So, we expect that this $l\approx 1^\circ$
enhancement in the microlensing rate is real and that it is related to the structure
and kinematics of the bulge.

\begin{figure}
\begin{center}
\includegraphics[angle=-90,scale=0.6,keepaspectratio]{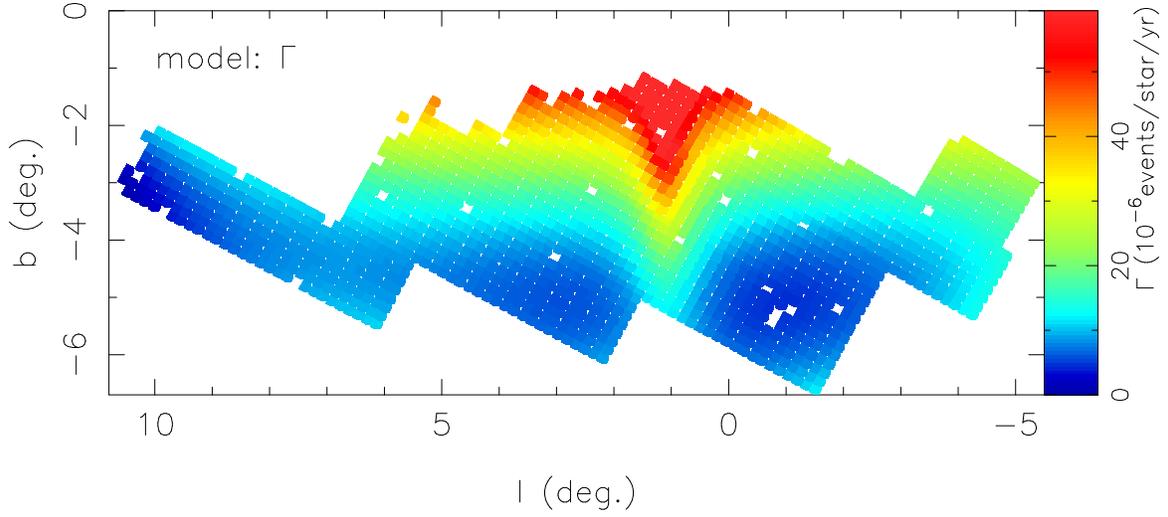}
\caption{
  \label{fig:Gamma_2D}
A 16-parameter model of microlensing event rate per star for the all-source sample. The 
model is described by Equation (\ref{eq:2Dformula}) with parameters given in Table~\ref{tbl:param_2D}.
}
\end{center}
\end{figure}

\begin{figure}
\begin{center}
\includegraphics[angle=-90,scale=0.6,keepaspectratio]{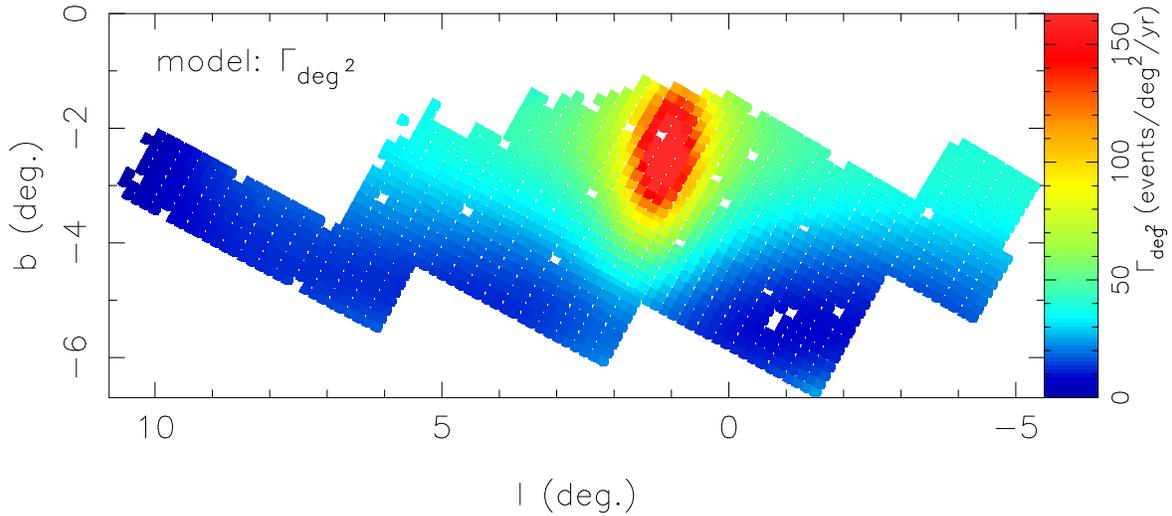}
\caption{
  \label{fig:Gammasq_2D}
A 16-parameter model of microlensing event rate per square degree for the all-source sample
with $I_s \leq 20$ mag.  The 
model is described by Equation (\ref{eq:2Dformula}) with parameters given in Table~\ref{tbl:param_2D}.
}
\end{center}
\end{figure}

Because of the relatively low noise in the $\Gamma$ and $\Gamma_{\rm deg^2}$ 
measurements, we have fit them with a 16-parameter model in $l$ and $b$. The
16 parameters consist of a 10-parameter cubic polynomial and the inverse of
a 6-parameter quadratic polynomial. That is
\begin{equation}
  \label{eq:2Dformula}
\begin{eqalign}
\Gamma = \ &
a_0  + 
a_1 l +
a_2 b+ 
a_3 l^2+
a_4 lb+
a_5 b^2+
a_6 l^3+
a_7 l^2b+
a_8 lb^2+
a_9  b^3 \\
& + 1/(
a_{10}  + 
a_{11} l +
a_{12} b+ 
a_{13} l^2+
a_{14} lb+
a_{15} b^2
) \ .
\end{eqalign}
\end{equation}
The best fit models for $\Gamma$ and $\Gamma_{\rm deg^2}$ are shown in Figure \ref{fig:Gamma_2D}  
and  \ref{fig:Gammasq_2D} and the model parameters are listed in Table~\ref{tbl:param_2D}.
Both models show the maximum at $l\approx 1^\circ$ that was also
evident in Figure \ref{fig:fieldmaps}.

\section{Discussion and conclusions}
\label{sec:discussionAndSummary}

\label{sec:summary}

We have measured the microlensing event rate and optical depth toward the 
Galactic bulge from the first two years of the MOA-II survey. 
Our sample of 474 events, with well measured parameters is larger than
all of the previous samples combined, and we employ a more careful treatment
of blending than all previous RCG samples. 
For the first time, we analyze the event rate and optical depth from for 
a sample of faint (often unresolved) stars and a RCG sample of 83 events
from the same data set. We are able to shed some light on the previously
noted difference between the optical depth measured with RCG samples
\citep{pop05,sumi2006,ham06} and samples of faint stars from DIA
photometry surveys \citep{alc00b,sumi03}. The faint star analyses
have shown systematically larger $\tau$ values.

Some of the previous RCG analyses \citep{pop05,ham06} have implicitly assumed that 
all events occurring in the proximity of RCG stars actually had
RCG sources, and then argued that blending caused relatively large errors
of both signs, which happened to nearly cancel in the measurement of $\tau$.
The OGLE-II analysis \citep{sumi2006} allowed for blending in the microlensing
light curve fits and was, like this analysis, restricted to events
with relatively small light curve parameter uncertainties. This allowed events
to be selected based on the best fit source brightness, which removes most
systematic errors due to blending. However, the photometry used by \citep{sumi2006} 
implicitly assumed that there was no blending, as it was done at the location of
the apparent star in the reference frame. In this analysis, we have performed the
photometry at the locations of the events as identified in difference images. As
a result, we avoid a systematic photometry error for blended events that might
affect the \citet{sumi2006} analysis, and our analysis method should be 
considered to be an improvement over the methods used for
previous RCG optical depth measurements.





\begin{figure*}
\begin{center}
\includegraphics[angle=-90,scale=0.65,keepaspectratio]{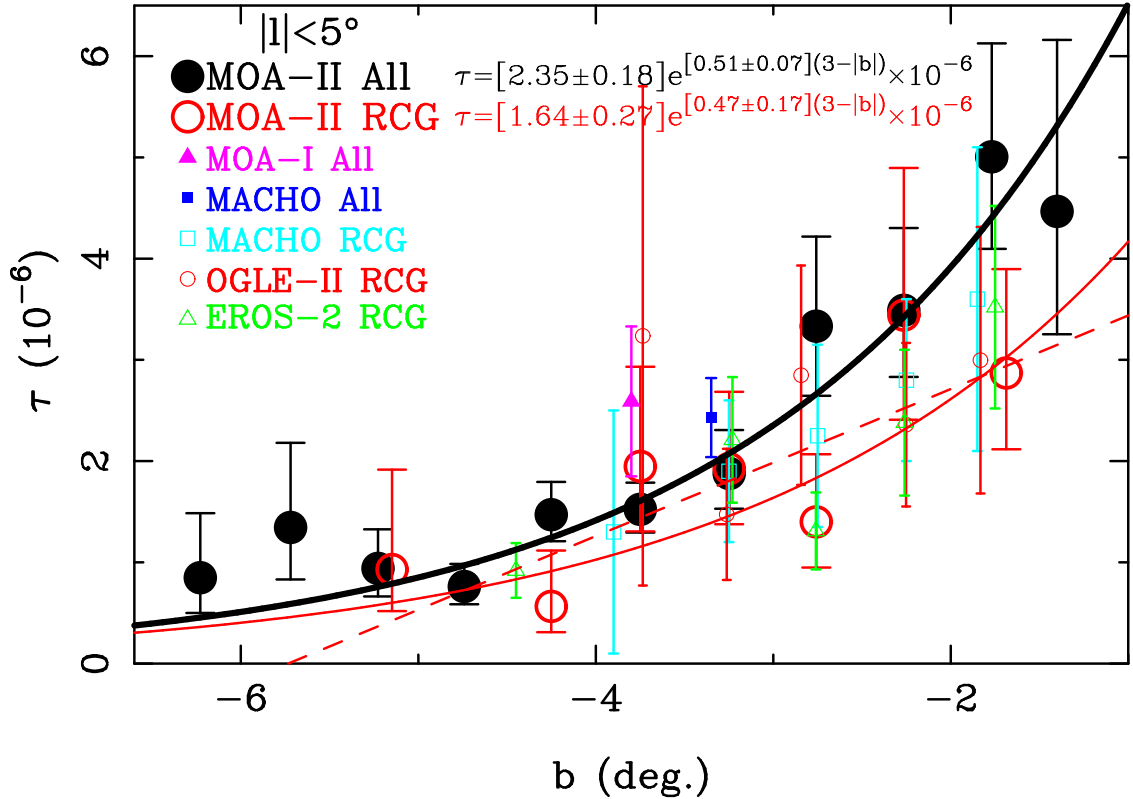}
\caption{The measured optical depth for the all-source (black filled circle) and 
RCG (red large open circle) samples as a function of galactic latitude $b$ for $|l|<5^\circ$. 
The subfields are combined into bins of width $\Delta b= 0.5^{\circ}$. The binned values are 
listed in Table \ref{tbl:opt_binb_all_lth5} and \ref{tbl:opt_binb_RCG_lth5}.
The filled circles, triangles and squares indicate $\tau$
for all-source samples measured by MOA-II (this work), MOA-I and MACHO surveys, respectively.
The red circles, open squares, circles and triangles denote the $\tau$ for RCG samples by the
MOA-II (this work), MACHO, OGLE-II and EROS surveys, respectively.
The thick black and thin red solid lines indicate the best fit exponential 
functions for the MOA-II measurements. The red dashed line denote the best linear model
for the OGLE-II RCG sample as a comparison.
\label{fig:optball} 
}
\end{center}
\end{figure*}

Our new all-source optical depth measurements are consistent will all 
previous measurements for both all-source and RCG samples, with the
exception of the EROS \citep{ham06}, which is more than 2-$\sigma$ 
smaller. Our RCG optical depth results are consistent with EROS and the
other previous RCG measurements, but we believe that our RCG optical depth
may be biased low, due to the low efficiency for long duration RCG events.
We can use the exponential models shown in Figure \ref{fig:optball} to interpolate
our measurement to the center of previous samples. For the MACHO DIA 
all-source result at $b = -3^\circ .35$ \citep{alc00b}, we find 
$\tau = [1.97\pm 0.15] \times 10^{-6}$ which
is 1.1-$\sigma$ smaller than the MACHO result of 
$\tau = 2.43{+0.39\atop -0.38}\times 10^{-6}$, using the
combined error bar. The MOA-I all-source result \citep{sumi03}, centered at 
$b = -3^\circ .8$, is $\tau = 2.59{+0.84\atop -0.64}\times 10^{-6}$. This compares to
our interpolated value of $\tau = [1.57\pm 0.12]\times 10^{-6}$, which is 1.4-$\sigma$ smaller.
Thus, our
new optical depth measurement is smaller at the $\sim 1$-$\sigma$ level than both 
the previous all-source measurements, which suggests that some of the
previously seen difference between the all-source and RCG samples is
due to statistical fluctuations.

The only previous RCG sample that distinguished RCG source events from 
events with main sequence sources that happened to be blended with
RCG stars was the OGLE-II analysis of \citet{sumi2006}, and they found
$\tau = 2.55{+0.57\atop -0.46}\times 10^{-6}$ at $b = -2^\circ .75$. This
compares to our all-source result, interpolated from the model given in 
Figure \ref{fig:optball}, is $\tau = [2.67\pm 0.20] \times 10^{-6}$, which is
just 0.2-$\sigma$ larger. Our RCG result, interpolated from the RCG
model in the same figure is
$\tau = [1.84\pm 0.30] \times 10^{-6}$, which is just 1.1-$\sigma$ smaller. 
The MACHO
Collaboration published several averages of their results \citep{pop05}, but
we compare to their ``CGR+3" average of $6\,{\rm deg}^2$ centered at
$b = -2^\circ.73$. MACHO reports $\tau = 2.37{+0.47\atop -0.39}\times 10^{-6}$
for RCG sources at this position. This compares to our interpolated 
all-source value of $\tau = [2.70\pm 0.21] \times 10^{-6}$, which is
just 0.7-$\sigma$ larger. Our interpolated RCG value at this position is
of $\tau = [1.86\pm 0.31] \times 10^{-6}$, which is 1.0-$\sigma$ smaller. 

The RCG sample
of the EROS Collaboration \citep{ham06} covers a wider area than the other
previous optical depth measurement samples, and in fact, it covers a
slightly larger area than the MOA-II analysis that we present here. Fortunately,
they fit their results to an exponential model that is identical to the one
shown in Figure \ref{fig:optball}, and they find
$\tau = [1.62\pm 0.23]\times10^{-6} \exp[(0.43\pm 0.16)(3-|b|)]$. This has a
slope that is consistent with our fits, which are shown in Figure \ref{fig:optball},
so we can compare our result to theirs by simply comparing the
normalization parameter of our models to theirs. For the all-source sample,
our normalization parameter is $[2.35\pm 0.18 ]\times10^{-6} $, which is 
2.4-$\sigma$ larger than the EROS value of $[1.62\pm 0.23 ]\times10^{-6} $.
A somewhat more fair comparison would be to compare to the the EROS
fit to a model fit to all our fields, instead of just those with $|l| < 5^\circ$.
This gives $\tau_{200} = [2.22\pm 0.16]\times10^{-6} \exp[(0.52\pm 0.07)(3-|b|)]$,
with is 2.1-$\sigma$ larger than the EROS value
However, our RCG value of $[1.64\pm 0.27 ]\times10^{-6} $ matches
their parameter, $[1.62\pm 0.23 ]\times10^{-6} $, to better than 0.1-$\sigma$. 
 
In summary, we find that our all-source results are about 1-$\sigma$ smaller
than the previous all-source measurements, and they are within 1-$\sigma$ 
of the RCG optical depth values from OGLE and MACHO. The only apparent
discrepancy is with the EROS measurement, which is just over 2-$\sigma$
smaller than our all-source optical depth. Partly, this is because the EROS
RCG sample is larger than the OGLE and MACHO ones, so the error bars
are smaller. It may also be that the different spatial coverage of the EROS
and MOA-II surveys plays a role in this difference. In any case, the odds
of one 2-$\sigma$ outlier out of 5 comparisons are about 25\%, so it is
fair to say that our all-source optical depth results are consistent with the
results for previous all-source and RCG samples. 

In any case, we find that our all-source optical depth measurement splits the 
range of previous measurements, coming in $\sim 1$-$\sigma$ below the 
previous all-source measurements and matching the MACHO and OGLE
RCG measurements to better than 1-$\sigma$
we find agreement within 1-$\sigma$
for 2 out of 5 comparisons and within 2-$\sigma$ for all 5. So, our results
are in good agreement with previous measurements. Our RCG optical depth
values also match the previous RCG measurements, but as we have
explain below, we believe that our RCG results are biased by poor
sensitivity to long timescale events.

Our optical depth results agree with many of the models that have been published.
Both the \citet{han03} and \cite{wood05} models agree with our all-source optical depth
to better than 1.3-$\sigma$ at $b = -3^\circ.9$ and $b = -3^\circ$, respectively,
but the \citet{wood05} model predicts an optical depth lower than the observed
value by 2.8-$\sigma$ or 23\% at $b = -2^\circ$. \citet{eva02} present a number of models, and their
``Dwek plus spiral structure" model agrees with our all-source optical depth
to 0.3-$\sigma$, while their other 
models predict both higher and lower optical depths. The model of \citet{bis02} predicted
a lower optical depth than previous measurements.

The theoretical modeling paper that comes closest to explaining our results is
\citet{kerins2009}. They consider several different event selection cuts: events
with a baseline magnitude $I < 19$, events with a peak magnitude $I < 19$,
and events with ``standard candle" sources, which are meant to correspond to the
RCG measurements. By comparing the contour levels of their optical depth maps,
we find that their baseline magnitude $I < 19$ results seem to match our measurements
well. Their $\tau = 4\times 10^{-6}$ contour is at $b = -1^\circ.9$, where the fit
to our measurements predicts $\tau = [4.12\pm 0.35]\times 10^{-6}$, and their
$\tau = 2\times 10^{-6}$ contour at $b = -3^\circ.5$, where the fit to our measurements
gives $\tau = [1.82\pm 0.15]\times 10^{-6}$. Our RCG sample gives $\tau$ values
that are 30-40\% below their predictions, but as we explain below, this is likely to be
due to the dearth of long timescale events in out RCG sample.

\begin{figure}
\begin{center}
\includegraphics[angle=-90,scale=0.45,keepaspectratio]{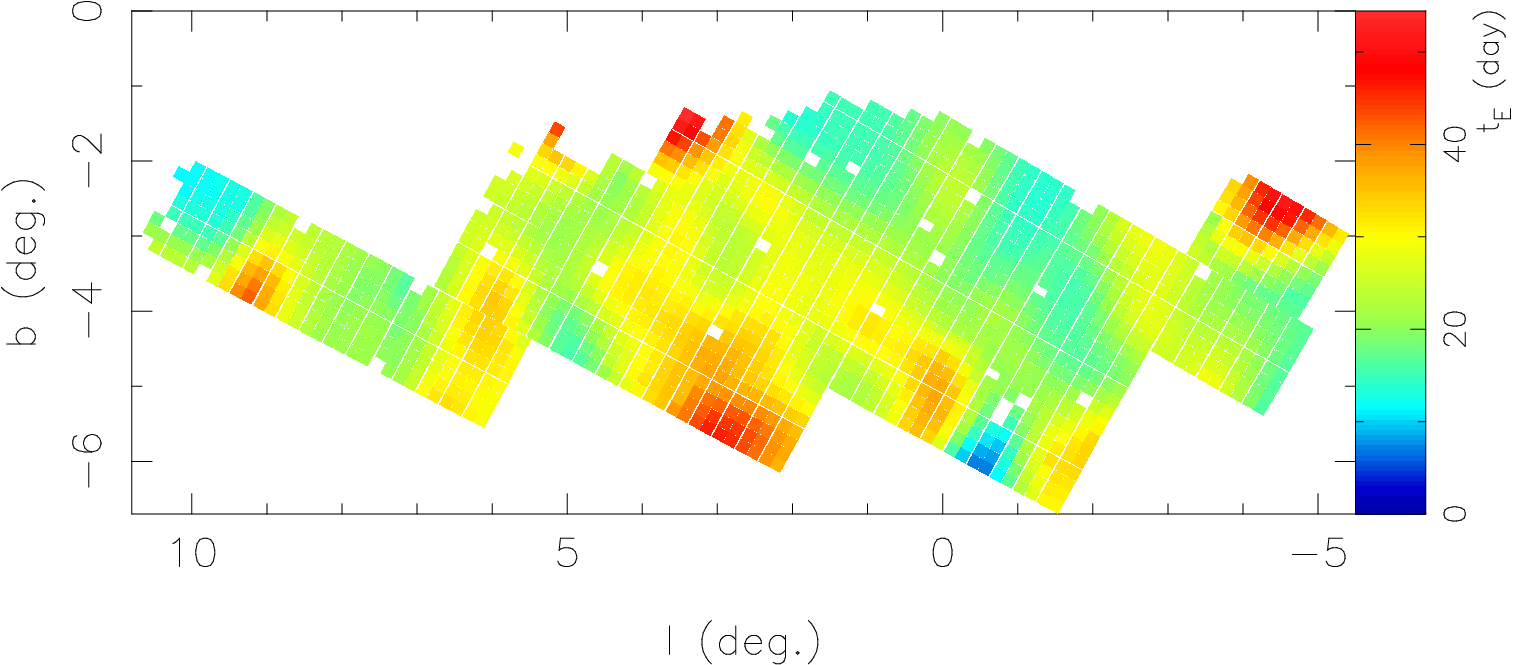}
\includegraphics[angle=-90,scale=0.45,keepaspectratio]{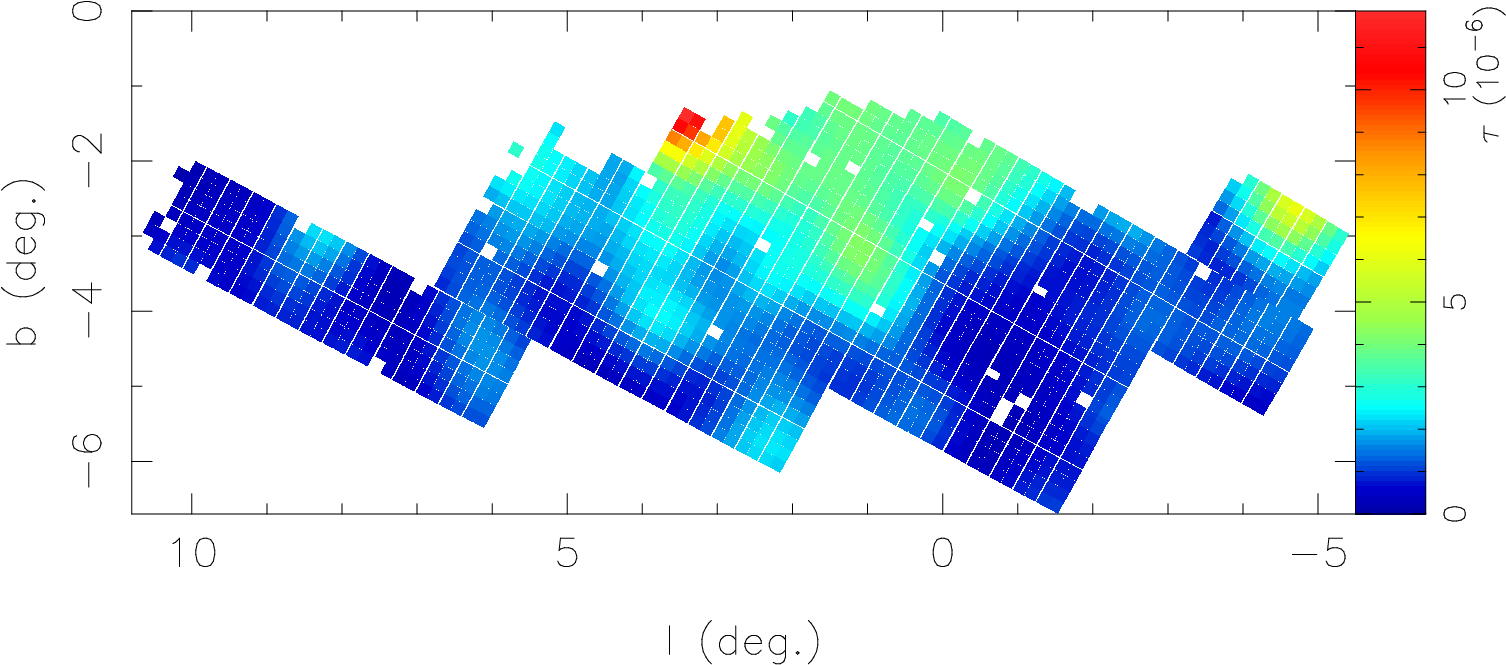}
\includegraphics[angle=-90,scale=0.45,keepaspectratio]{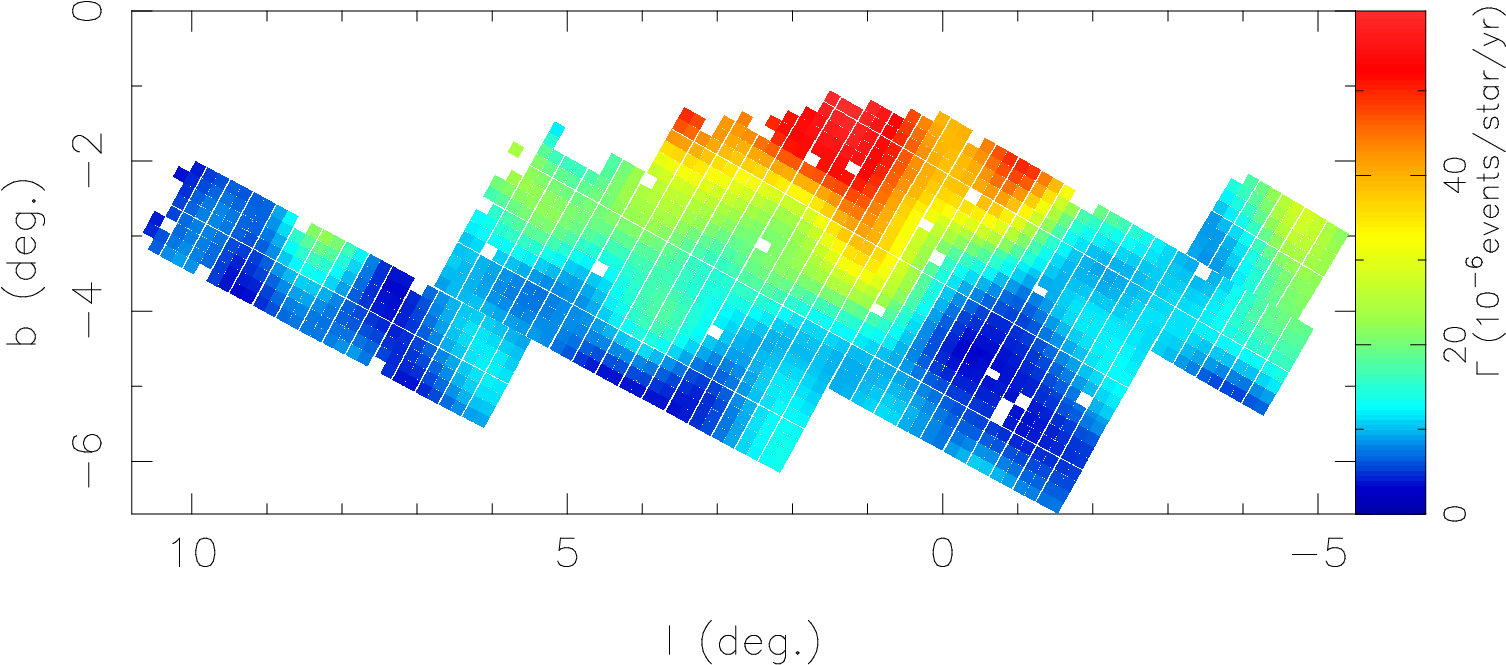}
\includegraphics[angle=-90,scale=0.45,keepaspectratio]{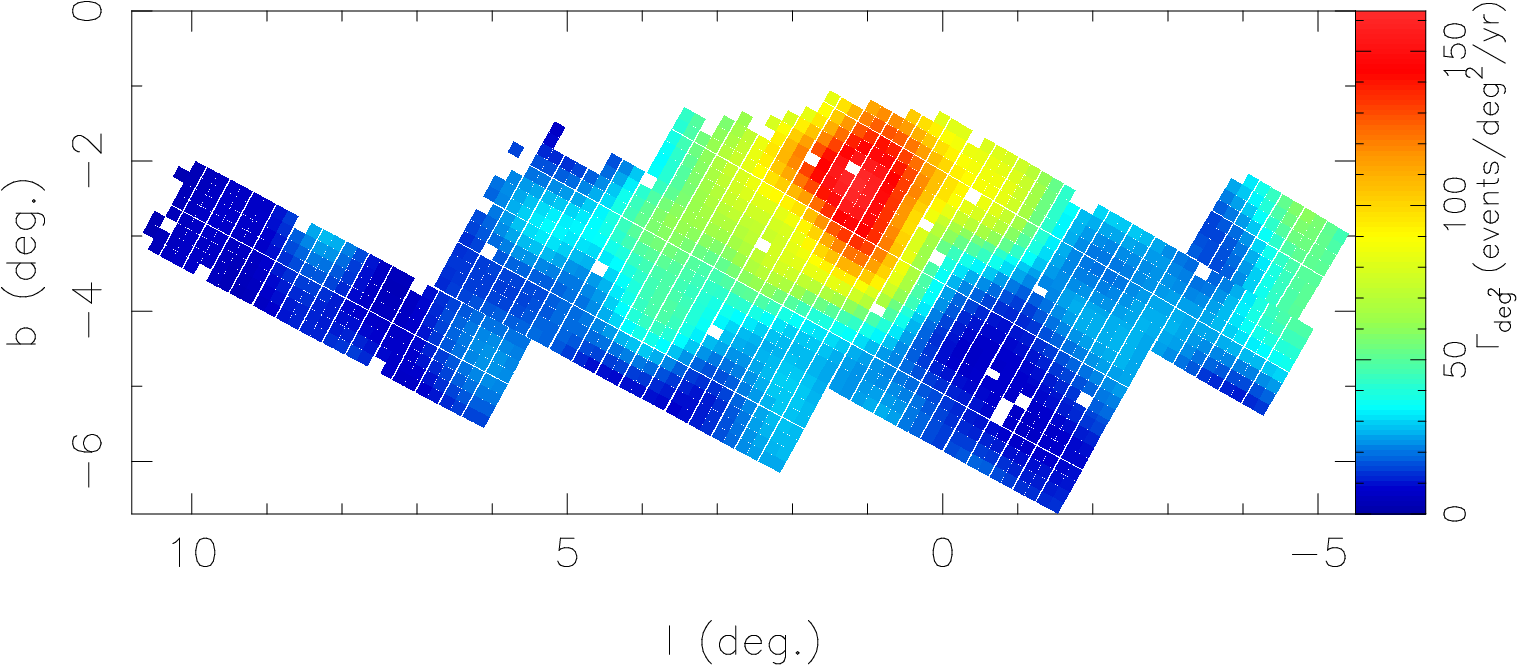}
\caption{
  \label{fig:fieldmaps}
False color (gray scale in the printed version) maps of the mean event timescale, $\VEV{t_{\rm E}}$, the 
measured optical depth, $\tau_{200}$, the event rate per star per year, $\Gamma$,  and the event rate per square 
degree per year, $\Gamma_{\rm deg^2}$, from top to bottom.
}
\end{center}
\end{figure}

\begin{figure*}
\begin{center}
\includegraphics[angle=-90,scale=0.65,keepaspectratio]{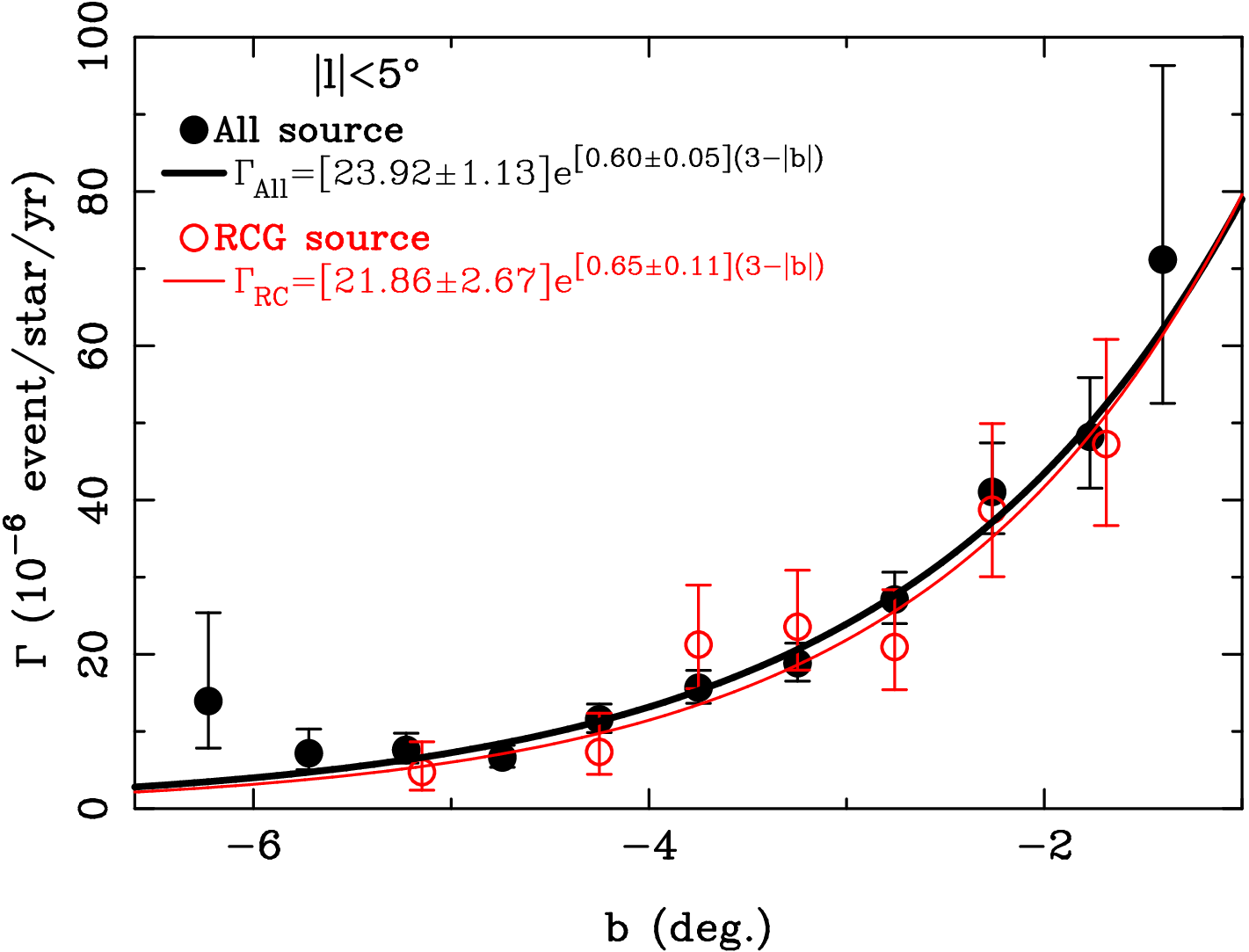}
\caption{The event rate per star per year, $\Gamma$, for the all-source (black filled circle) and RCG (red open circle) samples 
as a function of the galactic latitude $b$  for $|l|<5^\circ$,. 
The subfields are combined into bins of width $\Delta b= 0.5^{\circ}$ for display purposes
only, as the fitting was done using the unbinned subfield data with the Poisson statistics
fitting method.
The plotted values are listed in Tables \ref{tbl:opt_binb_all_lth5}
and \ref{tbl:opt_binb_RCG_lth5}.
The thick black and thin red solid lines indicate the best fit exponential functions 
for the all-source and RCG samples, respectively.
\label{fig:Gamma_vs_b} 
}
\end{center}
\end{figure*}

\begin{figure*}
\begin{center}
\includegraphics[angle=-90,scale=0.65,keepaspectratio]{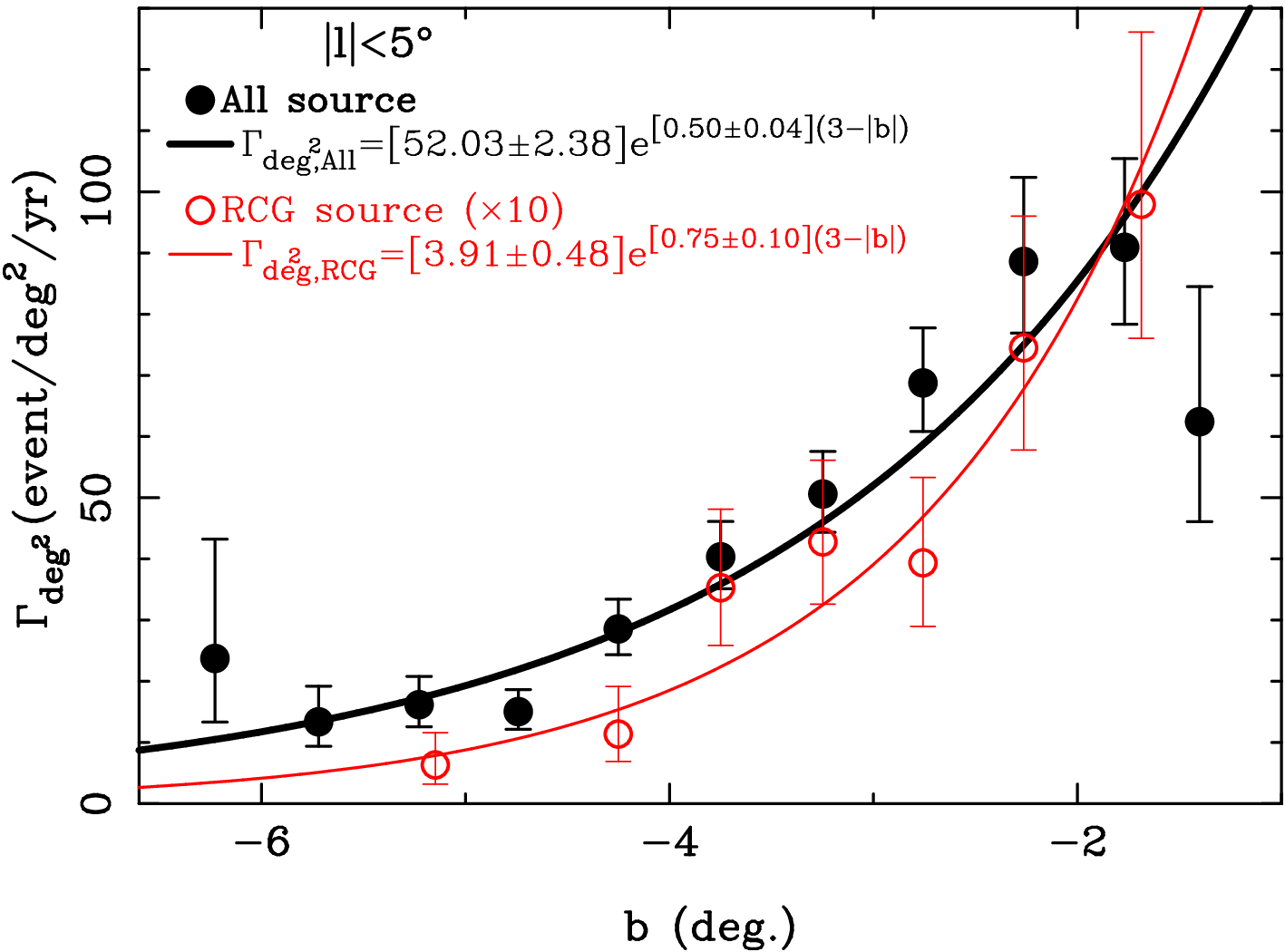}
\caption{The event rate per square degree per year, $\Gamma_{\rm deg^2}$, 
for the all-source (black filled circle) and RCG (red open circle) samples as a function of the galactic latitude $b$ for  $|l|<5^\circ$. 
The subfields are combined into bins of width $\Delta b= 0.5^{\circ}$ for display purposes
only, as the fitting was done using the unbinned subfield data with the Poisson statistics
fitting method.
The plotted values are listed in Tables \ref{tbl:opt_binb_all_lth5} and \ref{tbl:opt_binb_RCG_lth5}.
The thick black and thin red solid lines indicate the best fit exponential functions for 
all sources and RCG sample, respectively.
\label{fig:Gamma_sq_vs_b} 
}
\end{center}
\end{figure*}

Unfortunately, there are no theoretical predictions for the event rates
per star, $\Gamma$, and per square degree, $\Gamma_{\rm deg^2}$. 
As discussed above Section~\ref{sec:model}
and shown in Figures \ref{fig:fieldmaps}, \ref{fig:Gamma_vs_b}, and
\ref{fig:Gamma_sq_vs_b}, $\Gamma$ and $\Gamma_{\rm deg^2}$ can be
measured more precisely than $\tau$. Furthermore, $\tau$ has an additional
systematic uncertainty due to potential very long time scale events, which may
contribute significantly to $\tau$ but not to $\Gamma$ and $\Gamma_{\rm deg^2}$.



Figure \ref{fig:Gamma_vs_b} also indicates that the all-source and RCG $\Gamma$ values
differ by only $\sim 9$\%, which is much closer than the $\tau$ values at low latitudes.
While the all-source measurements should generally be more robust than the
RCG measurements, there is one weak point in all-source analysis. We don't directly
measure the number of sources. Instead we extrapolate to faint magnitudes using
the luminosity function of \citet{hol98}. It is reasonable to expect that the ratio of 
RCG stars to fainter main sequence stars could vary from the \citet{hol98}
value by 10-20\% at low latitudes, where most of the microlensing events are found.
If there are more faint stars than predicted by the  \citet{hol98} luminosity function,
there would be more microlensing events and a larger $\Gamma$ would be inferred.

\begin{figure}
\begin{center}
\includegraphics[angle=-90,scale=0.6,keepaspectratio]{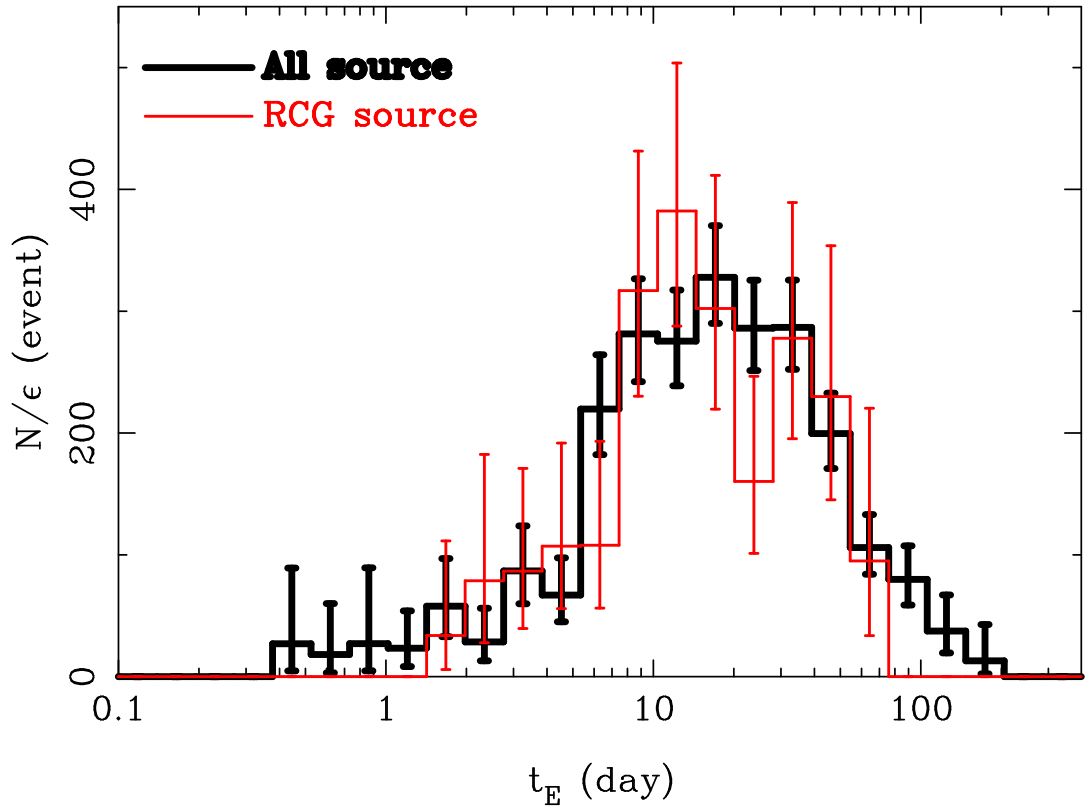}
\caption{
  \label{fig:tEdistribution}
The efficiency corrected Einstein radius crossing time distribution for the all-source 
sample in thick black and the RCG sample in thin red.
The RCG $t_{\rm E}$ distribution has been scaled to mach the amplitude of the all-source sample.
}
\end{center}
\end{figure}

\begin{figure}
\begin{center}
\includegraphics[angle=-90,scale=0.6,keepaspectratio]{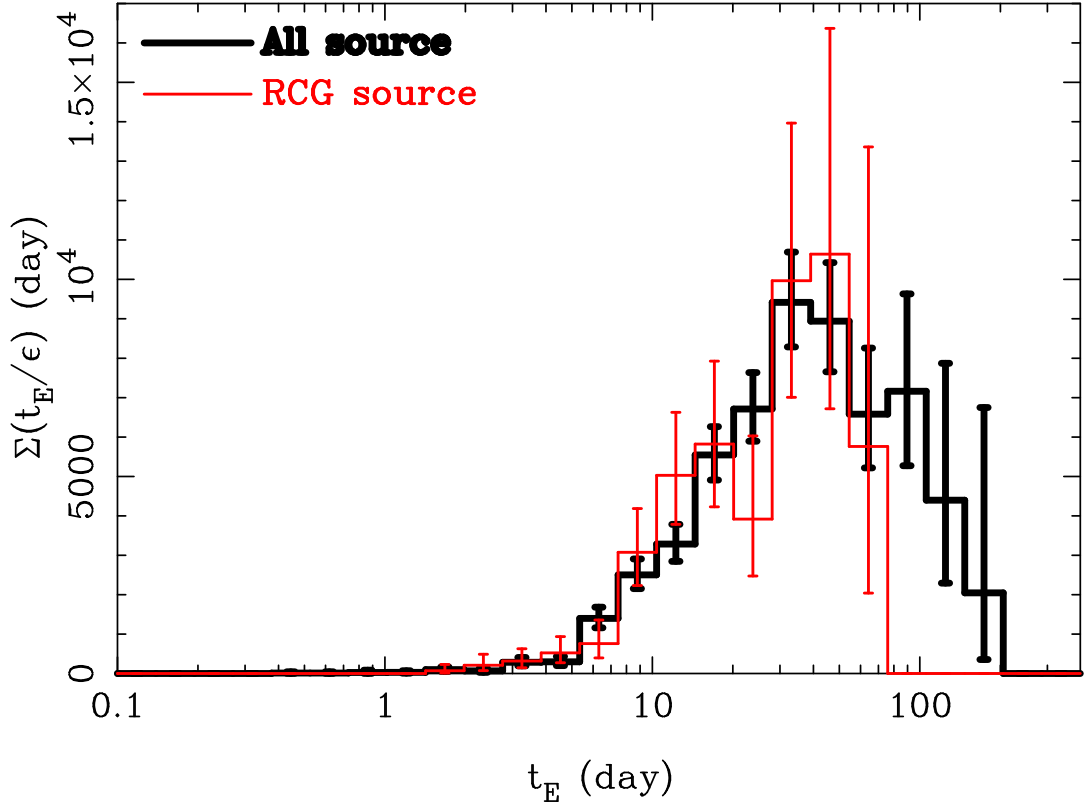}
\caption{
  \label{fig:tEdistribution2}
The efficiency corrected Einstein radius crossing time distribution, weighted by the 
$t_{\rm E}$ values, for the all-source sample in thick black and RCG sample in thin red.
This histogram indicates the contribution to the microlensing optical depth from each
$t_{\rm E}$ bin.
The amplitude of the RCG distribution has been scaled to mach the 
amplitude for the all-source distribution.
}
\end{center}
\end{figure}

\begin{figure}
\begin{center}
\includegraphics[angle=-90,scale=0.6,keepaspectratio]{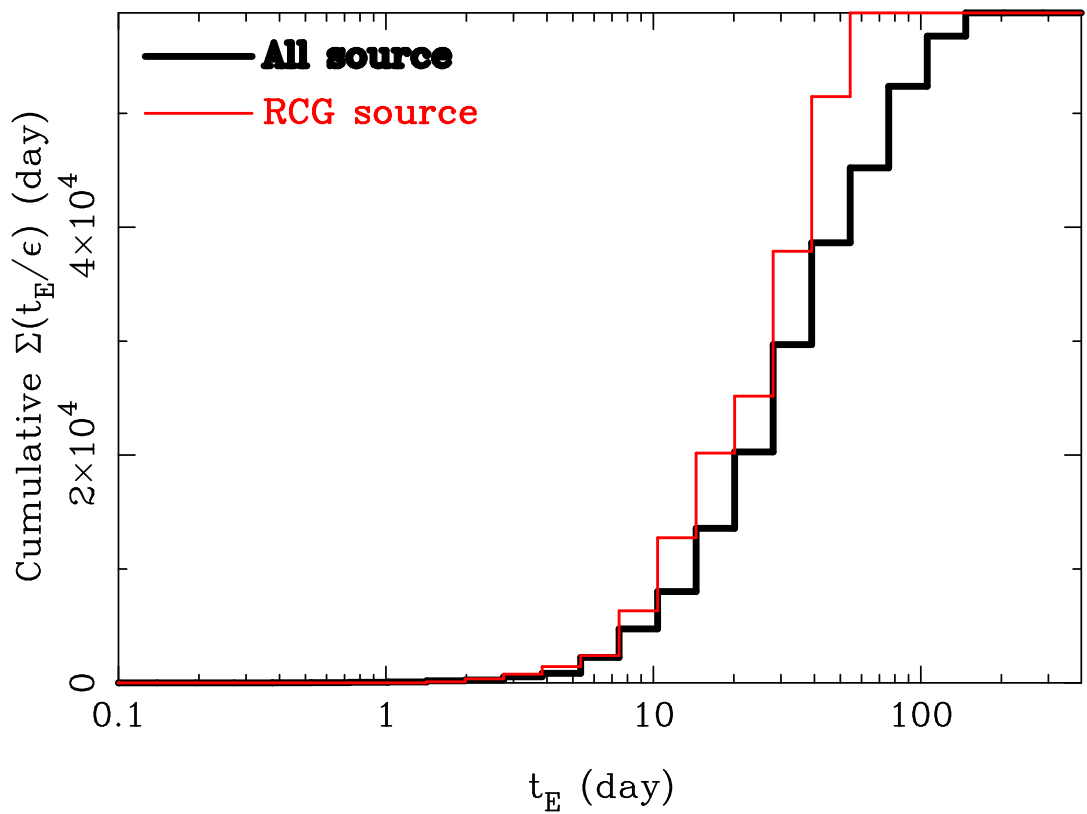}
\caption{
  \label{fig:tEdistribution3}
The cumulative distribution of efficiency corrected Einstein radius crossing time, weighted by the 
$t_{\rm E}$ values, for the all-source sample in thick black and RCG sample in thin red.
I.e.,  cumulative distribution of Figure \ref{fig:tEdistribution2}. This indicates the contribution to the 
microlensing optical depth from each $t_{\rm E}$ bin.
The amplitude of the RCG distribution has been scaled to mach the 
amplitude for the all-source distribution.
}
\end{center}
\end{figure}

Another significant difference between the all-source and RCG samples is their
(efficiency corrected) $t_{\rm E}$ distributions, which are shown in 
Figure \ref{fig:tEdistribution}. The mean timescale of the
all-source sample is larger than the mean timescale of the RCG sample.
Since $\tau_{200}\propto \VEV{t_{\rm E}}$, 
this is the reason why the all-source $\tau_{200}$ value is larger than the 
RCG $\tau_{200}$ value, while the $\Gamma$ values are consistent 
with each other. This may be largely due to the low detection efficiency 
at $t_{\rm E}>100$ days for the bright sources in the RCG sample, which is caused by
our requirement that the light curve be well fit by a constant brightness model
outside of a 120-day window centered on the event peak. Long duration
events with bright sources will deviate significantly from a constant baseline 
brightness at relatively low magnification, but fainter sources can pass this cut
because their photometric error bars are larger. For the all-source sample, events
with $t_E>80\,$days contribute 21\% to the
measurement of $\tau_{200}$, but the RCG sample has no events with 
$t_{\rm E} > 80\,$days. This is illustrated in Figure~\ref{fig:tEdistribution2}, which
shows the $t_{\rm E}$-weighted, efficiency corrected, timescale distributions for both
the all-source and RCG samples. The cumulative histogram of Figure~\ref{fig:tEdistribution2}
is also shown in Figure~\ref{fig:tEdistribution3}.
This is proportional to the contribution to $\tau_{200}$
for each $t_{\rm E}$ bin, and it is clear that the handful of events with $t_{\rm E} > 80\,$days
contribute significantly to $\tau_{200}$. This issue will be addressed in a future
analysis with a much longer time baseline. It is also possible that
the low latitude fields have an excess of
faint stars due to contamination from stars on the far side of the Galactic disk,
that might also contribute the larger $t_{\rm E}$ and $\tau_{200}$ values for the 
all-source sample.

It is our goal to measure $\Gamma$ and $\tau$ in various directions around the galactic bulge 
to constrain the model parameters of the barred Galactic bulge. Currently MOA-II detects
about 700 event per year and OGLE-IV finds about 1700 events a year. In the near future,
we plan to expand this analysis to include thousands of events which have been 
observed since the end of the 2007 observing season, which is the last season included
in this paper.

\begin{figure*}
\begin{center}
\includegraphics[angle=-90,scale=0.65,keepaspectratio]{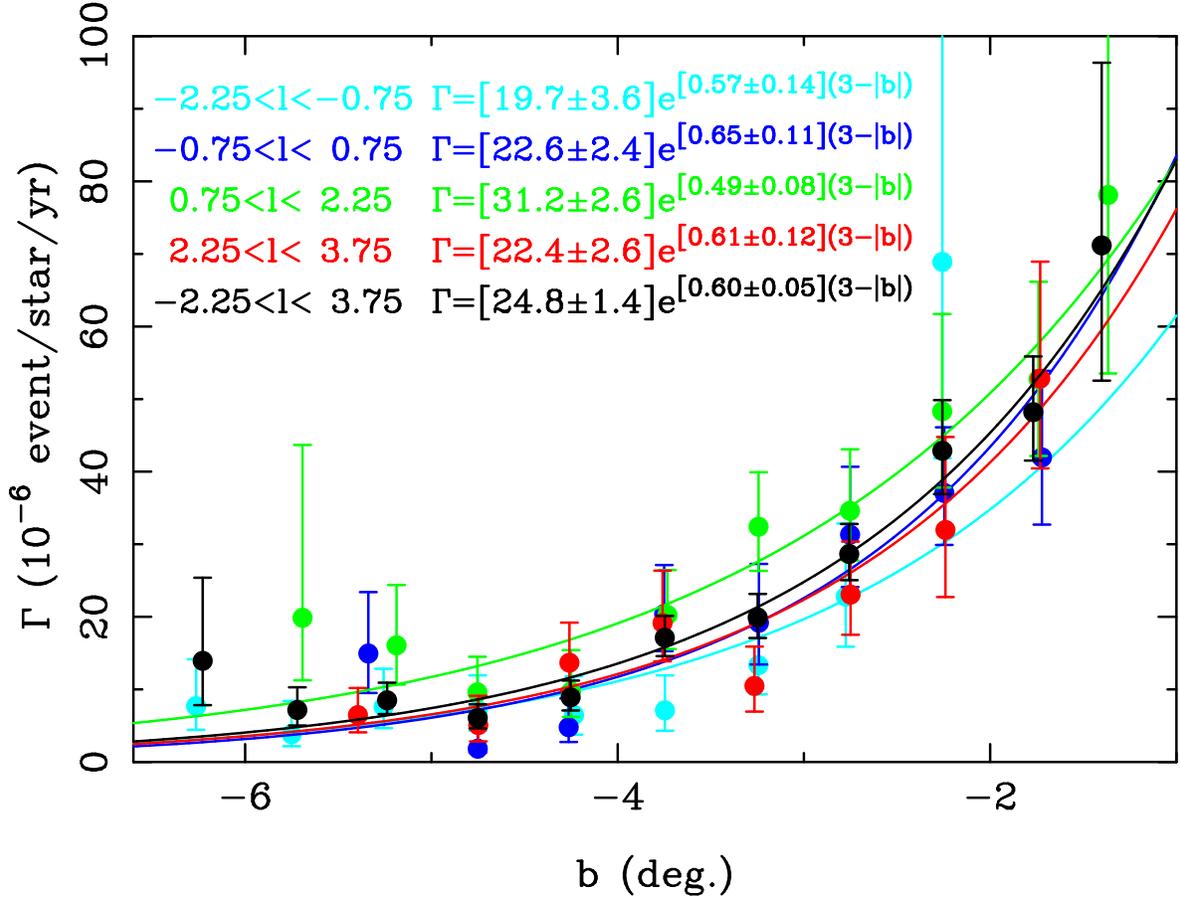}
\caption{The event rate per star per year, $\Gamma$, for the all-source sample 
as a function of the galactic latitude $b$ for different bins in Galactic
longitude, $l$. The curves show the best exponential fit in $b$ to the
unbinned subfield data. The black curve is the fit to all the events with
$-2^\circ.25 < l < 3^\circ.75$, and it provides a reasonable fit to all the
longitude bins, except the $0^\circ.75 < l < 2^\circ.25$ bin,
where there is an enhancement to the rate.
\label{fig:Gamma_vs_b_l} 
}
\end{center}
\end{figure*}

Another goal of this work is to predict the event rate in the inner 
Galactic bulge for the future infrared space microlensing survey of the 
Wide Field Infrared Space Telescope (WFIRST) \citep{green2012}, which was the
top ranked large space mission in the New Worlds, New Horizons (NWNH) 2010 decadal survey.
The exoplanet microlensing survey is one of the four major science programs
called out in NWNH, and one of two programs designated to drive the mission
design, along with the dark energy program. Our results can also be useful to 
the space exoplanet microlensing survey by Euclid \citep{penny13}.
The expected microlensing event rate for the WFIRST mission is uncertain because the region
with the highest event rate at the low galactic latitudes, where observations have
been sparse. This is partly due to the relatively small area of sky covered by some of
the previous microlensing surveys \citep{sumi2006}, but also because some of the 
previous surveys were designed for LMC observations and did not use a very
red passband \citep{alc00b,pop05,ham06}. The MOA-II survey improves this
situation somewhat due to its very wide field and custom red passband that
covers combined wavelength range of both the Cousins $R$ and $I$-bands,
but an infrared microlensing survey would be preferable. Nevertheless, the
measurements we present in this paper do provide the best estimate of the
event rate in the inner Galactic bulge to date. For $3.2\,{\rm deg}^2$ of
the MOA-II survey area inside $|b| \leq 3^\circ.0$ and $0^\circ.0 \leq l \leq 2^\circ.0$,
centered at  $(l,b)=(0.^\circ97, -2.^\circ26)$, we find 
$\Gamma =  4.57_{ -0.46}^{+  0.51} \times 10^{-5} {\rm yr}^{-1}$ for sources
with $I < 20$. This is a factor of 1.3 larger than the rate model used for the 
report of the WFIRST Science Definition Team (SDT) \citep{green2012}
evaluated at this position. However, the WFIRST SDT used a different model
to extrapolate to the lower latitude fields, $|b| \sim -1^\circ.4$, that WFIRST will 
observe. So, the model with $-2^\circ.25<l< 3^\circ.75$ presented in Figure~\ref{fig:Gamma_vs_b_l} predicts
a 60\% higher event rate than assumed in the report of the WFIRST SDT
\citep{green2012}.









\acknowledgments
We are grateful to B.S.\ Gaudi and M.\ Penny for helpful comments.
This MOA project is supported by the grant JSPS18253002 and JSPS20340052.
TS acknowledges the financial support from the JSPS, JSPS23340044, JSPS24253004. 
DPB acknowledges support from NSF grants AST-1009621 and AST-1211875,
as well as NASA grants NNX12AF54G and NNX13AF64G.


\clearpage





\begin{deluxetable}{lrrrrrrrrrc}
\tabletypesize{\scriptsize}
\tablecaption{MOA-II Galactic bulge fields with Galactic coordinates of the mean field center ($<l>$, $<b>$), 
the number of subfields used ($N_{\rm sub}$), the number of frames ($N_{\rm f}$), 
the number of source stars ($N_{\rm s}$ in thousands),
the number of microlensing events ($N_{\rm ev}$), 
the microlensing event rate per star per year ($\Gamma$),
the microlensing event rate per square degree per year ($\Gamma_{\rm deg^2}$), 
the optical depth ($\tau_{200}$), 
and the mean detection efficiency weighted $t_{\rm E}$. \label{tbl:fld}}
\tablewidth{0pt}
\tablehead{
    \colhead{Field} & \colhead{$<l>$} & \colhead{$<b>$} & \colhead{$N_{\rm sub}$}  
  & \colhead{$N_{\rm f}$}  & \colhead{$N_{\rm s}$} & \colhead{$N_{\rm ev}$} 
  & \colhead{$\tau_{200}$} & \colhead{$\Gamma (10^{-6})$} &  \colhead{$\Gamma_{\rm deg^2}$} & \colhead{$<t_{\rm E}>$} \\
    \colhead{} & \colhead{($^\circ$)} & \colhead{($^\circ$)} & \colhead{}  
  & \colhead{} & \colhead{$(10^3)$}  & \colhead{}
  & \colhead{(10$^{-6}$)} & \colhead{(star$^{-1}$yr$^{-1}$)} &  \colhead{(deg.$^{-2}$yr$^{-1}$)}  & \colhead{(day)}
}
\startdata
 gb1 & -4.3306 & -3.1119 &   79 & 2253 &    4240 &  22 &   2.56$_{ -0.87}^{+  1.93}$ & 19.0$_{-4.2}^{+5.3}$ & 37.0$_{-8.1}^{+10.4}$ & 30.6 \\
 gb2 & -3.8624 & -4.3936 &   79 & 2386 &    4741 &  17 &   0.94$_{ -0.22}^{+  0.29}$ & 11.3$_{-2.5}^{+3.2}$ & 24.7$_{-5.4}^{+6.9}$ & 18.8 \\
 gb3 & -2.3463 & -3.5133 &   79 & 2067 &    4589 &  19 &   1.31$_{ -0.31}^{+  0.42}$ & 11.1$_{-2.3}^{+2.9}$ & 23.5$_{-4.8}^{+6.2}$ & 26.7 \\
 gb4 & -0.8210 & -2.6317 &   77 & 2985 &    4239 &  41 &   3.00$_{ -0.59}^{+  0.75}$ & 35.1$_{-5.9}^{+7.3}$ & 70.2$_{-11.9}^{+14.7}$ & 19.4 \\
 gb5 &  0.6544 & -1.8595 &   65 & 8229 &    4457 &  67 &   3.85$_{ -0.53}^{+  0.61}$ & 50.1$_{-6.1}^{+6.7}$ & 124.9$_{-15.1}^{+16.7}$ & 17.4 \\
 gb6 &  1.8405 & -1.4890 &   11 & 1779 &     317 &   4 &   1.94$_{ -0.76}^{+  1.32}$ & 26.0$_{-10.4}^{+17.4}$ & 27.3$_{-10.9}^{+18.2}$ & 16.9 \\
 gb7 & -1.7147 & -4.5992 &   78 & 1970 &    4404 &  15 &   0.81$_{ -0.21}^{+  0.29}$ & 9.7$_{-2.3}^{+2.9}$ & 20.0$_{-4.6}^{+6.1}$ & 18.8 \\
 gb8 & -0.1937 & -3.7495 &   78 & 2139 &    5244 &  16 &   0.87$_{ -0.21}^{+  0.27}$ & 9.0$_{-2.0}^{+2.6}$ & 22.1$_{-4.9}^{+6.3}$ & 21.7 \\
 gb9 &  1.3329 & -2.8786 &   79 & 8301 &    7690 &  74 &   3.33$_{ -0.50}^{+  0.60}$ & 34.0$_{-4.0}^{+4.5}$ & 120.4$_{-14.0}^{+15.9}$ & 22.2 \\
gb10 &  2.8448 & -2.0903 &   70 & 1992 &    3707 &  36 &   4.84$_{ -1.23}^{+  1.65}$ & 38.8$_{-6.9}^{+8.6}$ & 74.6$_{-13.3}^{+16.6}$ & 28.3 \\
gb11 & -1.1093 & -5.7257 &   76 & 2004 &    3728 &   8 &   0.47$_{ -0.15}^{+  0.22}$ & 6.3$_{-2.1}^{+3.2}$ & 11.3$_{-3.8}^{+5.7}$ & 16.9 \\
gb12 &  0.4391 & -4.8658 &   79 & 1790 &    4861 &  12 &   1.06$_{ -0.34}^{+  0.49}$ & 7.2$_{-1.8}^{+2.4}$ & 16.2$_{-4.1}^{+5.4}$ & 33.2 \\
gb13 &  1.9751 & -4.0190 &   79 & 1811 &    6793 &  27 &   2.11$_{ -0.58}^{+  0.77}$ & 16.2$_{-3.0}^{+3.7}$ & 50.7$_{-9.3}^{+11.5}$ & 29.6 \\
gb14 &  3.5083 & -3.1698 &   79 & 1770 &    6304 &  29 &   1.98$_{ -0.53}^{+  0.72}$ & 17.6$_{-3.0}^{+3.7}$ & 51.2$_{-8.8}^{+10.8}$ & 25.5 \\
gb15 &  4.9940 & -2.4496 &   62 & 1952 &    1872 &  14 &   2.32$_{ -0.61}^{+  0.83}$ & 22.4$_{-5.3}^{+7.0}$ & 24.6$_{-5.8}^{+7.6}$ & 23.5 \\
gb16 &  2.6048 & -5.1681 &   79 & 1756 &    5043 &  17 &   1.56$_{ -0.42}^{+  0.56}$ & 10.0$_{-2.2}^{+2.8}$ & 23.3$_{-5.1}^{+6.5}$ & 35.3 \\
gb17 &  4.1498 & -4.3365 &   79 & 1792 &    5513 &  16 &   1.34$_{ -0.35}^{+  0.48}$ & 9.6$_{-2.2}^{+2.8}$ & 24.3$_{-5.5}^{+7.1}$ & 31.7 \\
gb18 &  5.6867 & -3.5055 &   78 & 1799 &    3841 &  13 &   1.01$_{ -0.29}^{+  0.41}$ & 9.5$_{-2.4}^{+3.1}$ & 17.1$_{-4.3}^{+5.6}$ & 23.9 \\
gb19 &  6.5534 & -4.5749 &   78 & 1704 &    3838 &  12 &   1.08$_{ -0.30}^{+  0.42}$ & 8.1$_{-2.0}^{+2.7}$ & 14.6$_{-3.7}^{+4.9}$ & 30.0 \\
gb20 &  8.1025 & -3.7531 &   79 & 1679 &    3037 &  12 &   1.17$_{ -0.32}^{+  0.44}$ & 10.8$_{-2.7}^{+3.6}$ & 15.1$_{-3.8}^{+5.0}$ & 24.6 \\
gb21 &  9.6172 & -2.9318 &   73 & 1659 &    1896 &   3 &   0.34$_{ -0.16}^{+  0.26}$ & 4.5$_{-2.1}^{+3.3}$ & 4.2$_{-2.0}^{+3.1}$ & 17.0 \\
all     &  1.8530 & -3.6890 & 1536 & --- &   90366 & 474 &   1.87$_{ -0.13}^{+  0.15}$ & 17.7$_{-0.9}^{+0.9}$ & 37.8$_{-1.9}^{+1.9}$ & 24.0 \\
all$_{\rm RC}^*$ &  1.8530 & -3.6890 & 1536 & --- &    6485 &  83 &   1.58$_{ -0.23}^{+  0.27}$ & 18.7$_{-2.0}^{+2.2}$ & 2.9$_{-0.3}^{+0.3}$ & 19.2 \\
\enddata
\tablecomments{
The values are for the all-source sample except for all$_{\rm RCG}$ which is for the RCG source sample.}
\end{deluxetable}

\clearpage

\begin{deluxetable}{lccrrrr}
\tabletypesize{\footnotesize}
\tablecaption{MOA-II subfields with Galactic coordinates, the numbers of source stars and events, 
and the mean timescale and efficiencies for the all-source sample.
 \label{tbl:sub_eff_all}}
\tablewidth{0pt}
\tablehead{
\colhead{subfield} &
\colhead{$l$} &
\colhead{$b$} &
\colhead{$N_{\rm s}$} &
\colhead{$N_{\rm ev}$} &
\colhead{$<t_{\rm E}>$} &
\colhead{$<\varepsilon>$} \\
\colhead{} &
\colhead{$(^\circ)$} &
\colhead{$(^\circ)$} &
\colhead{} &
\colhead{} &
\colhead{(days)} &
\colhead{} 
}
\startdata
   gb5-1-3 &  1.1704 & -1.3459 &    24032 &  2 & 15.4 & 0.2054 \\
   gb5-1-7 &  1.3125 & -1.2630 &    25504 &  1 & 15.4 & 0.2058 \\
   gb5-2-2 &  0.7835 & -1.3776 &    29267 &  1 & 15.5 & 0.1911 \\
   gb5-2-3 &  0.8685 & -1.5224 &    72017 &  0 & 15.4 & 0.1902 \\
   gb5-2-6 &  0.9280 & -1.2935 &    20980 &  0 & 15.2 & 0.1905 \\
   gb5-2-7 &  1.0130 & -1.4379 &    61885 &  1 & 15.2 & 0.1902 \\
   gb5-3-1 &  0.3942 & -1.4104 &    23886 &  0 & 17.6 & 0.1981 \\
   gb5-3-2 &  0.4788 & -1.5549 &    31046 &  0 & 17.4 & 0.1968 \\
   gb5-3-3 &  0.5639 & -1.6998 &    66763 &  1 & 16.8 & 0.1951 \\
   gb5-3-6 &  0.6239 & -1.4697 &    43697 &  2 & 16.3 & 0.1953 \\
   gb5-3-7 &  0.7091 & -1.6146 &    83169 &  1 & 15.9 & 0.1941 \\
   gb5-4-0 &  0.0089 & -1.4439 &    39811 &  0 & 19.8 & 0.1910 \\
   gb5-4-1 &  0.0918 & -1.5877 &    51173 &  0 & 20.3 & 0.1916 \\
   gb5-4-2 &  0.1755 & -1.7322 &    74557 &  1 & 20.7 & 0.1912 \\
   gb5-4-3 &  0.2599 & -1.8771 &    96247 &  1 & 20.3 & 0.1898 \\
   gb5-4-5 &  0.2356 & -1.5028 &    27844 &  0 & 19.1 & 0.1899 \\
   gb5-4-6 &  0.3197 & -1.6474 &    42434 &  2 & 18.9 & 0.1890 \\
   gb5-4-7 &  0.4044 & -1.7925 &    56646 &  3 & 18.6 & 0.1874 \\
   gb5-5-0 & -0.2872 & -1.6227 &    37610 &  1 & 20.4 & 0.1910 \\
   gb5-5-1 & -0.2055 & -1.7661 &    39203 &  0 & 21.9 & 0.1932 \\
   gb5-5-2 & -0.1227 & -1.9100 &    82900 &  0 & 22.8 & 0.1939 \\
\enddata
\tablecomments{
$N_{\rm s}$ and $N_{\rm ev}$ indicate the numbers of stars and microlensing events down to $I_{\rm s}<20$ mag.
The mean timescales, $<t_{\rm E}>$, are averaged over subfields within $1^\circ$ from the 
center of each subfield using a Gaussian weighting function with $\sigma = 0^\circ.4$.
$<\varepsilon>$ is average detection efficiency for the subfields by using the $t_{\rm E}$ distribution 
in the  the subfields within $1^\circ$ from the center of each subfield.
A complete electronic version of the table is available at http://iral2.ess.sci.osaka-u.ac.jp/\~{}sumi/OPTMOAII/Table.tar.gz.
}
\end{deluxetable}

\begin{deluxetable}{lccrrrr}
\tabletypesize{\footnotesize}
\tablecaption{MOA-II subfields with Galactic coordinates, the numbers of source stars and events, 
and the mean timescale and efficiencies for the RCG sample.
 \label{tbl:sub_eff_RCG}}
\tablewidth{0pt}
\tablehead{
\colhead{subfield} &
\colhead{$l$} &
\colhead{$b$} &
\colhead{$N_{\rm s}$} &
\colhead{$N_{\rm ev}$} &
\colhead{$<t_{\rm E}>$} &
\colhead{$<\varepsilon>$} \\
\colhead{} &
\colhead{$(^\circ)$} &
\colhead{$(^\circ)$} &
\colhead{} &
\colhead{} &
\colhead{(days)} &
\colhead{} 
}
\startdata
   gb5-1-3 &  1.1704 & -1.3459 &     5267 &  1 & 14.2 & 0.3363 \\
   gb5-1-7 &  1.3125 & -1.2630 &     5419 &  0 & 13.9 & 0.3365 \\
   gb5-2-2 &  0.7835 & -1.3776 &     5630 &  1 & 15.0 & 0.3673 \\
   gb5-2-3 &  0.8685 & -1.5224 &     8184 &  0 & 15.1 & 0.3683 \\
   gb5-2-6 &  0.9280 & -1.2935 &     4748 &  0 & 14.5 & 0.3670 \\
   gb5-2-7 &  1.0130 & -1.4379 &     7803 &  0 & 14.6 & 0.3684 \\
   gb5-3-1 &  0.3942 & -1.4104 &     4769 &  0 & 16.3 & 0.3965 \\
   gb5-3-2 &  0.4788 & -1.5549 &     5869 &  0 & 16.2 & 0.3974 \\
   gb5-3-3 &  0.5639 & -1.6998 &     7452 &  0 & 16.3 & 0.3984 \\
   gb5-3-6 &  0.6239 & -1.4697 &     6826 &  0 & 15.6 & 0.3974 \\
   gb5-3-7 &  0.7091 & -1.6146 &     7526 &  0 & 15.7 & 0.3986 \\
   gb5-4-0 &  0.0089 & -1.4439 &     6872 &  0 & 17.6 & 0.3784 \\
   gb5-4-1 &  0.0918 & -1.5877 &     7065 &  0 & 17.6 & 0.3784 \\
   gb5-4-2 &  0.1755 & -1.7322 &     7435 &  1 & 17.5 & 0.3787 \\
   gb5-4-3 &  0.2599 & -1.8771 &     7315 &  0 & 17.4 & 0.3790 \\
   gb5-4-5 &  0.2356 & -1.5028 &     5914 &  0 & 17.0 & 0.3785 \\
   gb5-4-6 &  0.3197 & -1.6474 &     6266 &  0 & 16.9 & 0.3788 \\
   gb5-4-7 &  0.4044 & -1.7925 &     7070 &  1 & 16.9 & 0.3792 \\
   gb5-5-0 & -0.2872 & -1.6227 &     6417 &  0 & 18.7 & 0.4131 \\
   gb5-5-1 & -0.2055 & -1.7661 &     6684 &  0 & 18.6 & 0.4131 \\
   gb5-5-2 & -0.1227 & -1.9100 &     7564 &  0 & 18.5 & 0.4133 \\
\enddata
\tablecomments{
Notation is the same as Table \ref{tbl:sub_eff_all}, but for the RCG sample.
A complete electronic version of the table is available  at http://iral2.ess.sci.osaka-u.ac.jp/\~{}sumi/OPTMOAII/Table.tar.gz.
}
\end{deluxetable}

\begin{deluxetable}{lccrrrlcc}
\tabletypesize{\scriptsize}
\tablecaption{Microlensing events used in the optical depth and event rate measurements.
\label{tbl:candlist}
}
\tablewidth{0pt}
\tablehead{
\colhead{ID} & 
\colhead{R.A.} & 
\colhead{Dec.} & 
\colhead{$N_{\rm data}$}& 
\colhead{$t_0$}  & 
\colhead{$t_{\rm E}$} & 
\colhead{$\ \ u_{\rm 0}$} & 
\colhead{$I_{\rm s}$} & 
\colhead{$\frac{\chi^2}{dof}$}  \\
\colhead{} &
\colhead{(2000)} &
\colhead{(2000)} &
\colhead{}&
\colhead{(JD$^\prime$)}  &
\colhead{(day)} &
\colhead{} &
\colhead{(mag)} &
\colhead{}  
}
\startdata
      gb1-R-1-14 & 17:45:11.248 & -33:38:54.84 &  2052 & 3828.94193 &  23.60$\pm$2.28 & 0.909150$\pm$0.144886 &  16.8 & 0.99 \\
   gb1-R-1-52516 & 17:45:46.378 & -33:36:05.22 &  2073 & 4031.68805 &  23.02$\pm$0.73 & 0.090750$\pm$0.003734 &  18.1 & 1.58 \\
   gb1-R-2-22735 & 17:45:35.960 & -33:45:56.56 &  2074 & 3883.90740 &  22.39$\pm$0.39 & 0.212003$\pm$0.005573 &  18.0 & 0.67 \\
   gb1-R-3-64176 & 17:44:52.898 & -34:21:36.52 &  2092 & 4193.71721 &  10.64$\pm$0.14 & 0.766208$\pm$0.015656 &  14.9 & 1.20 \\
          gb1-R-3-76 & 17:45:40.324 & -34:20:04.75 &  2106 & 3825.27275 & 157.57$\pm$11.00 & 0.188031$\pm$0.020548 &  19.7 & 0.87 \\
   gb1-R-3-46605 & 17:45:24.278 & -34:05:15.32 &  2088 & 4001.26676 &  16.60$\pm$1.18 & 0.173574$\pm$0.043284 &  18.6 & 0.89 \\
   gb1-R-4-47848 & 17:45:45.039 & -34:27:53.52 &  2103 & 4303.16456 &  14.98$\pm$1.14 & 0.331496$\pm$0.041186 &  19.5 & 0.84 \\
   gb1-R-4-18284 & 17:46:24.506 & -34:30:36.82 &  2101 & 3883.24171 &   0.73$\pm$0.08 & 0.028096$\pm$0.003360 &  19.7 & 0.77 \\
     gb1-R-4-910 & 17:46:53.887 & -34:30:42.59 &  2095 & 3887.62778 &  26.14$\pm$1.06 & 0.587581$\pm$0.038963 &  17.7 & 0.68 \\
   gb1-R-5-65454 & 17:45:22.188 & -34:57:04.75 &  2075 & 4221.40839 &  40.11$\pm$0.45 & 0.190260$\pm$0.002905 &  17.5 & 1.66 \\
   gb1-R-5-82191 & 17:46:01.446 & -34:54:35.01 &  2089 & 4312.71403 &  14.23$\pm$0.41 & 0.106859$\pm$0.004776 &  19.2 & 1.31 \\
   gb1-R-6-77689 & 17:50:28.442 & -34:49:23.09 &  2068 & 4364.59588 &  45.51$\pm$0.74 & 0.413738$\pm$0.010909 &  17.1 & 1.30 \\
   gb1-R-6-82519 & 17:50:05.385 & -34:51:36.71 &  2073 & 4362.33309 &   3.26$\pm$0.54 & 0.840722$\pm$0.290765 &  17.1 & 1.51 \\
   gb1-R-6-79294 & 17:49:50.655 & -34:53:49.40 &  2063 & 4340.06886 &  23.06$\pm$0.94 & 0.375246$\pm$0.023845 &  18.7 & 0.77 \\
   gb1-R-6-65555 & 17:47:53.567 & -34:50:55.15 &  2076 & 4241.08796 &   4.92$\pm$0.32 & 0.050412$\pm$0.004092 &  19.4 & 1.48 \\
   gb1-R-6-40898 & 17:49:46.072 & -35:05:13.00 &  2061 & 4019.80173 &  11.24$\pm$0.69 & 0.475032$\pm$0.049014 &  17.6 & 0.73 \\
   gb1-R-6-73986 & 17:48:11.363 & -35:02:34.80 &  2081 & 4291.58993 &  19.18$\pm$0.70 & 0.153344$\pm$0.007234 &  19.6 & 0.76 \\
   gb1-R-7-71783 & 17:50:17.617 & -34:27:00.09 &  2075 & 4303.67369 &  30.97$\pm$2.54 & 0.464778$\pm$0.054527 &  19.3 & 1.06 \\
   gb1-R-7-30409 & 17:47:53.653 & -34:29:04.26 &  2062 & 3953.51717 &  12.43$\pm$1.16 & 0.810060$\pm$0.140936 &  17.8 & 0.65 \\
\enddata
\tablecomments{The error bars for $t_{\rm E}$ and $u_0$ indicate 68\% confidence intervals.
JD$^\prime = {\rm JD}-2450000$.
$I_s$ indicates the best fit $I$-band source magnitude.
$\chi^2/dof$ is the reduced chi-squre for the best fit single-lens model.
The complete table is available electronically at  http://iral2.ess.sci.osaka-u.ac.jp/\~{}sumi/OPTMOAII/Table.tar.gz.  
The print edition contains only a sample.
\\}
\end{deluxetable}

\clearpage

\begin{deluxetable}{rrrrrrr}
\tabletypesize{\footnotesize}
\tablecaption{Microlensing optical depth and event rates binned in $b$ for the all-source
sample with $|l|<5^\circ$.
 \label{tbl:opt_binb_all_lth5}}
\tablewidth{0pt}
\tablehead{
\colhead{$<b>^*$} &
\colhead{$N_{\rm sub}$} &
\colhead{$N_{\rm s}$} &
\colhead{$N_{\rm ev}$} &
\colhead{$\tau (10^{-6})$} &
\colhead{$\Gamma$\,$(10^{-6})$} &
\colhead{$\Gamma_{\rm deg^2}$} \\
\colhead{$(^\circ)$} &
\colhead{$$} &
\colhead{} &
\colhead{$$} &
\colhead{$$} &
\colhead{(star$^{-1}$ yr$^{-1}$)} &
\colhead{(deg.$^{-2}$yr$^{-1}$)} 
}
\startdata
  -1.4012 &  20  &   482422 &  12 & 4.47$_{ -1.21}^{+1.69}$ &   71.2$_{-18.6}^{+ 25.2}$ &   62.4$_{-16.3}^{+ 22.1}$\\
  -1.7690 &  70  &  3631956 &  52 & 5.01$_{ -0.91}^{+1.12}$ &   48.2$_{ -6.7}^{+  7.7}$ &   90.9$_{-12.6}^{+ 14.5}$\\
  -2.2645 & 114  &  6766400 &  70 & 3.49$_{ -0.66}^{+0.81}$ &   41.1$_{ -5.4}^{+  6.4}$ &   88.6$_{-11.7}^{+ 13.7}$\\
  -2.7576 & 146  & 10190175 &  75 & 3.33$_{ -0.69}^{+0.88}$ &   27.1$_{ -3.1}^{+  3.5}$ &   68.8$_{ -7.9}^{+  9.0}$\\
  -3.2486 & 168  & 12407499 &  67 & 1.88$_{ -0.35}^{+0.43}$ &   18.8$_{ -2.3}^{+  2.6}$ &   50.6$_{ -6.3}^{+  7.0}$\\
  -3.7490 & 172  & 12182546 &  58 & 1.52$_{ -0.23}^{+0.26}$ &   15.7$_{ -2.0}^{+  2.2}$ &   40.3$_{ -5.2}^{+  5.8}$\\
  -4.2512 & 172  & 11677303 &  43 & 1.47$_{ -0.26}^{+0.32}$ &   11.6$_{ -1.7}^{+  2.0}$ &   28.6$_{ -4.2}^{+  4.9}$\\
  -4.7410 & 154  &  9620731 &  22 & 0.76$_{ -0.18}^{+0.22}$ &    6.6$_{ -1.3}^{+  1.6}$ &   15.0$_{ -2.9}^{+  3.6}$\\
  -5.2270 & 101  &  5911839 &  16 & 0.94$_{ -0.28}^{+0.39}$ &    7.6$_{ -1.7}^{+  2.2}$ &   16.2$_{ -3.6}^{+  4.6}$\\
  -5.7197 &  56  &  2874105 &   8 & 1.34$_{ -0.51}^{+0.84}$ &    7.2$_{ -2.1}^{+  3.1}$ &   13.4$_{ -4.0}^{+  5.8}$\\
  -6.2282 &  21  &   983156 &   4 & 0.85$_{ -0.35}^{+0.64}$ &   13.9$_{ -6.1}^{+ 11.5}$ &   23.7$_{-10.4}^{+ 19.5}$\\
\enddata
\tablecomments{
$*$Average galactic latitude of fields in each bin.
$N_{\rm sub}$, $N_{\rm s}$ and $N_{\rm ev}$ indicate the number of subfields, source stars and microlensing events
in each bin.
}
\end{deluxetable}
\begin{deluxetable}{rrrrrrrrr}
\tabletypesize{\footnotesize}
\tablecaption{Microlensing optical depth and event rates binned in $b$ for the RCG
sample with $|l|<5^\circ$.
 \label{tbl:opt_binb_RCG_lth5}}
\tablewidth{0pt}
\tablehead{
\colhead{$<b>^*$} &
\colhead{$N_{\rm sub}$} &
\colhead{$N_{\rm s}$} &
\colhead{$N_{\rm ev}$} &
\colhead{$\tau (10^{-6})$} &
\colhead{$\Gamma$\,$(10^{-6})$} &
\colhead{$\Gamma_{\rm deg^2}$} \\
\colhead{$(^\circ)$} &
\colhead{$$} &
\colhead{} &
\colhead{$$} &
\colhead{$$} &
\colhead{(star$^{-1}$ yr$^{-1}$)} &
\colhead{(deg.$^{-2}$yr$^{-1}$)} 
}
\startdata
  -1.6872 &  90  &   512987 &  16 & 2.87$_{ -0.75}^{+1.03}$ &   47.3$_{-10.6}^{+ 13.6}$ &    9.8$_{ -2.2}^{+  2.8}$\\
  -2.2645 & 114  &   602968 &  16 & 3.44$_{ -1.03}^{+1.45}$ &   38.7$_{ -8.7}^{+ 11.2}$ &    7.5$_{ -1.7}^{+  2.2}$\\
  -2.7576 & 146  &   754360 &  11 & 1.40$_{ -0.45}^{+0.67}$ &   20.9$_{ -5.5}^{+  7.4}$ &    3.9$_{ -1.0}^{+  1.4}$\\
  -3.2486 & 168  &   839277 &  14 & 1.93$_{ -0.55}^{+0.76}$ &   23.6$_{ -5.6}^{+  7.3}$ &    4.3$_{ -1.0}^{+  1.3}$\\
  -3.7490 & 172  &   785581 &  11 & 1.95$_{ -0.64}^{+0.98}$ &   21.2$_{ -5.7}^{+  7.7}$ &    3.5$_{ -0.9}^{+  1.3}$\\
  -4.2512 & 172  &   734458 &   4 & 0.57$_{ -0.26}^{+0.55}$ &    7.3$_{ -2.9}^{+  5.0}$ &    1.1$_{ -0.4}^{+  0.8}$\\
  -5.1480 & 332  &  1225445 &   3 & 0.93$_{ -0.41}^{+0.99}$ &    4.7$_{ -2.4}^{+  3.9}$ &    0.6$_{ -0.3}^{+  0.5}$\\
\enddata
\tablecomments{
$*$Average galactic latitude of fields in each bin. The notation is the same as in Table \ref{tbl:opt_binb_all_lth5}.
}
\end{deluxetable}

\begin{deluxetable}{lccrrrrrr}
\tabletypesize{\footnotesize}
\tablecaption{Average microlensing optical depth and event rates at the position of each
subfield for the all-source sample.
 \label{tbl:opt_2D}}
\tablewidth{0pt}
\tablehead{
\colhead{subfield} &
\colhead{$l$} &
\colhead{$b$} &
\colhead{$N_{\rm sub}$} &
\colhead{$N_{\rm s}$} &
\colhead{$N_{\rm ev}$} &
\colhead{$\tau (10^{-6})$} &
\colhead{$\Gamma$ ($10^{-6}$)} &
\colhead{$\Gamma_{\rm deg^2}$}  \\
\colhead{} &
\colhead{$(^\circ)$} &
\colhead{$(^\circ)$} &
\colhead{} &
\colhead{} &
\colhead{} &
\colhead{} &
\colhead{(star$^{-1}$yr$^{-1}$)} &
\colhead{(deg.$^{-2}$yr$^{-1})$} 
}
\startdata
   gb5-1-3 &  1.1704 & -1.3459 &  58 & 3476887 & 55 & $ 4.0_{ -0.8}^{+ 0.9}$ & $ 59.3_{-14.3}^{+18.0}$ & $109.0_{-26.3}^{+ 33.1}$\\
   gb5-1-7 &  1.3125 & -1.2630 &  51 & 2826886 & 46 & $ 4.0_{ -0.8}^{+ 1.0}$ & $ 59.4_{-15.3}^{+21.5}$ & $ 96.6_{-24.8}^{+ 34.9}$\\
   gb5-2-2 &  0.7835 & -1.3776 &  59 & 3756406 & 59 & $ 3.8_{ -0.6}^{+ 0.8}$ & $ 56.1_{-12.7}^{+16.4}$ & $118.0_{-26.7}^{+ 34.5}$\\
   gb5-2-3 &  0.8685 & -1.5224 &  69 & 4605788 & 67 & $ 3.8_{ -0.6}^{+ 0.7}$ & $ 55.8_{-11.0}^{+13.8}$ & $126.1_{-24.8}^{+ 31.1}$\\
   gb5-2-6 &  0.9280 & -1.2935 &  54 & 3210920 & 49 & $ 3.9_{ -0.7}^{+ 0.9}$ & $ 58.4_{-14.2}^{+18.7}$ & $112.9_{-27.4}^{+ 36.1}$\\
   gb5-2-7 &  1.0130 & -1.4379 &  63 & 4042440 & 63 & $ 3.9_{ -0.7}^{+ 0.8}$ & $ 58.1_{-12.2}^{+15.5}$ & $120.2_{-25.2}^{+ 32.1}$\\
   gb5-3-1 &  0.3942 & -1.4104 &  53 & 3639406 & 52 & $ 3.7_{ -0.7}^{+ 0.9}$ & $ 47.6_{-12.0}^{+15.8}$ & $103.0_{-26.1}^{+ 34.2}$\\
   gb5-3-2 &  0.4788 & -1.5549 &  67 & 4635780 & 63 & $ 3.7_{ -0.7}^{+ 0.8}$ & $ 48.5_{-10.1}^{+12.8}$ & $114.3_{-23.7}^{+ 30.2}$\\
   gb5-3-3 &  0.5639 & -1.6998 &  78 & 5615435 & 74 & $ 3.7_{ -0.6}^{+ 0.7}$ & $ 49.4_{ -9.4}^{+11.4}$ & $126.2_{-24.1}^{+ 29.1}$\\
   gb5-3-6 &  0.6239 & -1.4697 &  65 & 4326789 & 65 & $ 3.8_{ -0.6}^{+ 0.7}$ & $ 52.4_{-11.3}^{+14.6}$ & $117.9_{-25.4}^{+ 32.8}$\\
   gb5-3-7 &  0.7091 & -1.6146 &  76 & 5229492 & 75 & $ 3.7_{ -0.6}^{+ 0.7}$ & $ 53.0_{ -9.9}^{+12.4}$ & $128.6_{-24.0}^{+ 30.2}$\\
   gb5-4-0 &  0.0089 & -1.4439 &  48 & 3061436 & 37 & $ 3.5_{ -0.8}^{+ 1.0}$ & $ 39.9_{-11.8}^{+15.6}$ & $ 82.1_{-24.2}^{+ 32.2}$\\
   gb5-4-1 &  0.0918 & -1.5877 &  59 & 3949695 & 52 & $ 3.7_{ -0.8}^{+ 1.0}$ & $ 40.9_{ -9.7}^{+12.9}$ & $ 92.6_{-21.9}^{+ 29.2}$\\
   gb5-4-2 &  0.1755 & -1.7322 &  73 & 5041433 & 65 & $ 3.8_{ -0.7}^{+ 0.9}$ & $ 41.9_{ -8.9}^{+10.9}$ & $104.0_{-22.1}^{+ 27.0}$\\
   gb5-4-3 &  0.2599 & -1.8771 &  81 & 5740732 & 70 & $ 3.8_{ -0.7}^{+ 0.8}$ & $ 42.7_{ -8.0}^{+ 9.9}$ & $114.7_{-21.5}^{+ 26.5}$\\
   gb5-4-5 &  0.2356 & -1.5028 &  56 & 3878445 & 53 & $ 3.7_{ -0.7}^{+ 0.9}$ & $ 43.6_{-10.8}^{+13.9}$ & $ 97.5_{-24.2}^{+ 31.1}$\\
   gb5-4-6 &  0.3197 & -1.6474 &  70 & 4909533 & 67 & $ 3.8_{ -0.7}^{+ 0.8}$ & $ 45.1_{ -9.3}^{+11.8}$ & $110.1_{-22.8}^{+ 28.9}$\\
   gb5-4-7 &  0.4044 & -1.7925 &  81 & 5774820 & 76 & $ 3.8_{ -0.6}^{+ 0.7}$ & $ 46.1_{ -9.0}^{+10.8}$ & $121.9_{-23.9}^{+ 28.4}$\\
   gb5-5-0 & -0.2872 & -1.6227 &  53 & 3233763 & 43 & $ 3.6_{ -0.7}^{+ 0.9}$ & $ 40.1_{-12.3}^{+17.5}$ & $ 81.0_{-24.8}^{+ 35.3}$\\
   gb5-5-1 & -0.2055 & -1.7661 &  63 & 4002999 & 50 & $ 3.8_{ -0.8}^{+ 0.9}$ & $ 39.3_{-10.3}^{+13.9}$ & $ 86.8_{-22.9}^{+ 30.6}$\\
   gb5-5-2 & -0.1227 & -1.9100 &  76 & 4986751 & 57 & $ 3.9_{ -0.8}^{+ 1.0}$ & $ 38.7_{ -9.3}^{+12.3}$ & $ 93.1_{-22.4}^{+ 29.5}$\\
\enddata
\tablecomments{
The averages include all the subfields within $1^\circ$ of the center of each subfield with
a Gaussian weighting function with $\sigma = 0^\circ .4$.
$N_{\rm sub}$, $N_{\rm s}$ and $N_{\rm ev}$ are numbers of subfields, source stars and microlensing events 
in this $1^\circ$ circle, respectively. 
A complete electronic version of this table is available at http://iral2.ess.sci.osaka-u.ac.jp/\~{}sumi/OPTMOAII/Table.tar.gz
}
\end{deluxetable}
\begin{deluxetable}{lrr}
\tablecaption{The best 2D model parameters for  $\Gamma$ and $\Gamma_{\rm deg^2}$.
 \label{tbl:param_2D}}
\tablewidth{0pt}
\tablehead{
\colhead{param} &
\colhead{$\Gamma$} &
\colhead{$\Gamma_{\rm deg^2}$} \\
}
\startdata
$ a_{ 0} $ &    93.032844 &   -24.481156 \\
$ a_{ 1} $ &     1.248177 &     5.539010 \\
$ a_{ 2} $ &    36.846116 &   -73.451537 \\
$ a_{ 3} $ &    -0.282139 &    -0.790312 \\
$ a_{ 4} $ &     0.405687 &     3.113193 \\
$ a_{ 5} $ &     4.142922 &   -24.290347 \\
$ a_{ 6} $ &    -0.025380 &    -0.020730 \\
$ a_{ 7} $ &    -0.124668 &    -0.244327 \\
$ a_{ 8} $ &     0.021965 &     0.370164 \\
$ a_{ 9} $ &     0.072898 &    -2.156637 \\
$ a_{10} $ &     0.337819 &     0.055451 \\
$ a_{11} $ &    -0.496257 &    -0.037629 \\
$ a_{12} $ &     0.035555 &     0.024214 \\
$ a_{13} $ &     0.218693 &     0.027694 \\
$ a_{14} $ &    -0.004649 &     0.009795 \\
$ a_{15} $ &     0.007841 &     0.007221 \\
\enddata
\tablecomments{  The model parameters are defined in Equation (\ref{eq:2Dformula}).
}
\end{deluxetable}

\end{document}